\documentclass[manuscript,screen]{acmart}
\usepackage{tcolorbox}
\AtBeginDocument{%
  }

\setcopyright{acmlicensed}
\copyrightyear{2025}
\acmYear{2025}
\acmDOI{XXXXXXX.XXXXXXX}
\acmISBN{978-1-4503-XXXX-X/2018/06}

\usepackage{tcolorbox}
\usepackage{soul}
\usepackage{graphicx}
\usepackage{subcaption}
\usepackage{tabularx}
\usepackage{pifont}
\usepackage{multirow}
\usepackage{array}
\usepackage{longtable}
\usepackage{xcolor}

\begin{document}

\title{Dependability of UAV-Based Networks and Computing Systems: A Survey}

\author{Qingyang Zhang}
\orcid{0000-0002-0625-0428}
\affiliation{%
  \institution{University of Tsukuba}
  \city{Tsukuba}
  \state{Ibaraki}
  \country{Japan}}
\email{zhang.qingyang@sd.cs.tsukuba.ac.jp}

\author{Mohammad Dwipa Furqan}
\orcid{0009-0006-5794-0222}
\affiliation{%
  \institution{University of Tsukuba}
  \city{Tsukuba}
  \state{Ibaraki}
  \country{Japan}}
\email{dwipa.mohammad@sd.cs.tsukuba.ac.jp}

\author{Tasfia Nuzhat}
\orcid{0009-0008-4109-8469}
\affiliation{%
  \institution{University of Tsukuba}
  \city{Tsukuba}
  \state{Ibaraki}
  \country{Japan}}
\email{nuzhat.tasfia@sd.cs.tsukuba.ac.jp}

\author{Fumio Machida}
\orcid{0000-0001-9779-983X}
\affiliation{%
  \institution{University of Tsukuba}
  \city{Tsukuba}
    \state{Ibaraki}
  \country{Japan}}
\email{machida@cs.tsukuba.ac.jp}

\author{Ermeson Andrade}
\orcid{0000-0002-9614-4492}
\affiliation{%
  \institution{Federal Rural University of Pernambuco}
  \city{Recife}
  \country{Brazil}}
\email{ermeson.andrade@ufrpe.br}


\begin{abstract}
Uncrewed Aerial Vehicle (UAV) computing and networking are becoming a fundamental computation infrastructure for diverse cyber-physical application systems. UAVs can be empowered by AI on edge devices and can communicate with other UAVs and ground stations via wireless communication networks. Dynamic computation demands and heterogeneous computing resources are distributed in the system and need to be controlled to maintain the quality of services and to accomplish critical missions. With the evolution of UAV-based systems, dependability assurance of such systems emerges as a crucial challenge. UAV-based systems confront diverse sources of uncertainty that may threaten their dependability, such as software bugs, component failures, network disconnections, battery shortages, and disturbances from the real world. In this paper, we conduct systematic literature reviews on the dependability of UAV-based networks and computing systems. The survey report reveals emerging research trends in this field and summarizes the literature into comprehensive categories by threat types and adopted technologies. Based on our literature reviews, we identify eight research fields that require further exploration in the future to achieve dependable UAV-based systems.
\end{abstract}




\maketitle

\section{Introduction}
\label{sec:introduction}

Uncrewed Aerial Vehicles (UAVs), also known as drones, have rapidly evolved from niche tools to critical components in a wide array of cyber-physical systems (CPS), enabling applications such as disaster response \cite{Wei2022-oq}, precision agriculture \cite{Velez2024-gf}, surveillance \cite{Zhang2021-em}, and logistics \cite{Xu2024-wl}. 
By integrating advanced computing paradigms and capabilities at the edge and leveraging wireless communication networks within 5/6G connectivity \cite{De_Paula_Soares2023-ek}, UAVs can perform complex tasks autonomously or collaboratively, often in dynamic and unpredictable environments. 
Equipped with artificial intelligence (AI) and interconnected with other UAVs and ground stations \cite{Ning2024-vu}, these systems form a distributed computing and networking infrastructure that supports real-time decision-making and mission execution.
However, the increasing complexity, autonomy, and interconnectivity of UAV-based CPS present substantial challenges to their dependability. This concept encompasses reliability, availability, maintainability, and safety, all of which are critical to the secure and effective operation of these systems \cite{avizienis2004basic}.

Dependability has long been a cornerstone of cyber-physical systems \cite{miclea2011dependability}. 
Traditional CPS frameworks emphasize fault tolerance, redundancy, and rigorous verification to mitigate failures.
UAV-based systems, as a specialized class of CPS, inherit these principles but face unique challenges due to their aerial mobility, limited energy resources, and exposure to dynamic environmental conditions (e.g., wind gusts, electromagnetic interference) \cite{Pacheco2021-wy, Xue2023-qx, Zhan2024-dw}.
For instance, while ground-based CPS may rely on stable power sources and fixed communication infrastructure, UAVs must balance limited battery life with computational demands, all while maintaining connectivity in ad-hoc airborne networks. 
These fundamental differences require a re-examination of existing dependability paradigms to effectively address the specific vulnerabilities and operational demands of UAV-based systems.

Dependability in UAV-based systems encompasses a range of attributes, including reliability, availability, safety, and resilience, all of which are essential to maintaining Quality of Service (QoS) and accomplishing mission objectives \cite{Gupta2022-ax, Farrukh2023-lp, avizienis2004basic}. 
These systems operate in environments where they face diverse sources of uncertainty, such as hardware and software failures, network disruptions, energy constraints, environmental disturbances, and communication breakdowns. 
For instance, a UAV performing a search-and-rescue mission may encounter battery depletion or signal loss, potentially leading to mission failure or safety hazards \cite{Kaymaz2024-kg}. 
Recent studies have identified energy constraints as a primary threat, as battery depletion often causes mission aborts in long-duration operations \cite{Lin2021-an}. 
Morereover, network disruptions in UAV swarms can hinder coordination and data sharing due to intermittent connectivity \cite{Alioua2018-mv}.

Additionally, environmental factors like extreme weather exacerbate hardware and software failures, further complicating dependable operation.
As UAV-based systems become more integral to critical applications, ensuring their dependability has emerged as a pressing research challenge, necessitating a systematic understanding of the threats they face and the methodologies to mitigate them.

Efforts to improve dependability of UAV-based systems have led to a variety of proposed approaches, such as heuristic optimization techniques for task offloading to edge servers \cite{Baktayan2022-dj}, energy harvesting mechanisms to extend flight duration \cite{Kim2021-uy}, and fault-tolerant protocols for 5G-enabled UAV swarms to ensure reliable communication under disruptions \cite{Damigos2023-fa}. 
Simulation techniques, including Markov models, have also been employed to evaluate these solutions in dynamic environments \cite{Bai2023-cz, Zhao2024-ke}. 
Despite these advancements, the literature remains fragmented, with studies often focusing on isolated aspects of dependability, including energy management or network reliability, without integrating both networking and computing dimensions.

This survey report aims to provide a comprehensive overview of the dependability of UAV-based networks and computing systems, synthesizing the state-of-the-art research and identifying key trends, challenges, and opportunities. To achieve this, we conducted a Systematic Literature Review (SLR) following a rigorous methodology to ensure a thorough and unbiased selection of relevant studies. We first identified a set of keywords related to dependability of UAV-based systems and system type, such as "UAV reliability," "dependability", and "UAV computing system".
Using these keywords, we searched five major literature databases, including IEEE Xplore \footnote{https://ieeexplore.ieee.org/Xplore/home.jsp}, ACM Digital Library \footnote{https://dl.acm.org/}, SpringerLink \footnote{https://link.springer.com/}, ScienceDirect \footnote{https://www.sciencedirect.com/}, and ISI Web of Science \footnote{http://www.isiknowledge.com}, collecting publications from the past decade (January 2015 to July 2024). 
This initial search yielded a total of 1,848 articles. 
We then performed a meticulous screening process by reviewing the titles and abstracts of these articles, applying inclusion and exclusion criteria to focus on studies directly addressing the dependability of UAV-based systems in networking and computing contexts. 
After this filtering, we selected 458 articles for in-depth analysis, forming the foundation of this survey.

To guide our analysis, we formulated the following research questions: 
\begin{itemize}
    \item[(1)] 
         How have research trends on the dependability of UAV-based systems evolved in recent years?
    \item[(2)] 
         What types of dependability threats are considered in the literature on UAV-based networks and computing systems?
    \item[(3)] 
         What techniques are adopted to improve the dependability of UAV-based networks and computing systems?
    \item[(4)] 
         What are the future research directions for dependable UAV-based networks and computing systems?
\end{itemize}
These questions provide a structured framework to explore the multifaceted aspects of dependability of UAV-based systems, from identifying vulnerabilities to evaluating solutions and envisioning future advancements.

The rest of the survey is organized as follows.  We begin by reviewing related work in Section \ref{sec:related work} to contextualize the evolution of UAV-based systems and their dependability requirements. 
Section \ref{sec:methodology} introduces the methodology used to conduct the research.
We then explore research trends in Section \ref{sec:research trends}, including systematic taxonomies, categorization of dependability aspects, and comparative analyses of metrics, system types, threats, and methodological approaches. 
In Section \ref{sec:dependability threats}, a detailed examination of dependability threats
highlights the multifaceted nature of the problem. 
To address these challenges, we survey methodological approaches in Section \ref{sec:Dependability Techniques}.
Then, we discuss future research directions in Section \ref{sec:future research} and threats to validity in Section \ref{sec:threats to validity}, offering insights into the evolving landscape of dependability of UAV-based systems. 
Finally, we summarize our work in Section \ref{sec:conclusion}.


\section{Related work}
\label{sec:related work}

A growing body of surveys has explored various aspects of UAV-based computing and networking.

\paragraph{UAV-Enabled Edge and Distributed Computing.}
Recent surveys have focused on UAV-enabled edge and distributed computing systems.
For example, Xia et al. \cite{xia2023survey} presented a granular classification of resource management paradigms, such as computation offloading, energy-aware scheduling, and hybrid edge-cloud collaboration. 
Huda et al. \cite{huda2022survey} further compared model-based optimization and learning-driven methods in offloading decision-making frameworks, highlighting algorithmic trade-offs in responsiveness, scalability, and adaptability.
Adil et al. \cite{adil2024uav} examined UAV-assisted IoT applications, outlining QoS requirements and communication challenges that directly impact system reliability.
These works highlight the importance of efficient resource management and robust communication for dependable UAV operations, yet they often treat these aspects in isolation.

\paragraph{AI Techniques in UAV Systems.}
The intersection of AI and UAVs has also garnered significant attention.
Ning et al. \cite{ning2023mobile} mapped ML techniques (e.g., federated learning, reinforcement learning) to UAV applications like autonomous navigation and swarm coordination.
Heidari et al. \cite{heidari2023machine} explored the role of machine learning in enhancing UAV autonomy and decision-making, while also addressing energy efficiency through offloading techniques.
These surveys underscore the potential of AI to mitigate dependability challenges, but they rarely connect these advancements to broader dependability frameworks that encompass both networking and computing.

\paragraph{Energy and Sustainability Considerations.}
Sustainable operation remains a pressing challenge for UAV-based systems.
Ahmed et al. \cite{ahmed2025toward} emphasized energy sustainability, addressing optimization in power consumption and flight planning.
Cyrille et al. \cite{cyrille2024critical} categorized energy management strategies across hardware optimization and renewable energy harvesting.
They highlight the need for integrated energy solutions that balance computational and networking demands, an area that intersects with dependability but is often studied independently.

\paragraph{Security and Privacy in UAV Networks.}
Additionally, surveys on cyber-physical security and privacy, such as those discussing secure communication and data protection in UAV networks, highlight the intersection of dependability and security, a critical concern in safety-critical applications.
Zhi et al. \cite{zhi2020security} provided an overview of UAV security threats across sensors, communication links, and multi-UAV coordination, while also addressing privacy concerns such as unauthorized surveillance and data leakage.
Mekdad et al. \cite{mekdad2023survey} analyzed vulnerabilities in both civilian and military UAV applications, highlighting challenges in secure transmission and adversarial defense. 
These studies are complementary to our study, as security and privacy are also essential properties when systems are threatened by malicious attackers. In this paper, we limit our scope to dependability (i.e., reliability, availability, maintainability, and safety) and do not cover security and privacy-related issues.

\paragraph{Survey Gaps and Motivation.}
While these surveys provide valuable insights into specific topics of UAV computing and networking, they focus on isolated aspects, such as communication, computation, or security, without a holistic examination of dependability across both networking and computing domains. 
This fragmentation highlights the need for a comprehensive survey, such as the one proposed in this work, that systematically integrates these perspectives to address the multifaceted nature of dependability of UAV-based systems.

\section{Methodology}
\label{sec:methodology}

Our systematic mapping was guided by the methodology proposed by Petersen et al.\cite{petersen2008systematic}, which we employed to identify works related to UAV dependability under different computing systems and approaches.

\subsection{Planning the Review}


To ensure systematic coverage, we formulated a set of focused research questions (RQs) as follows.
\begin{itemize}
    \item[(1)]  
         How have research trends on the dependability of UAV-based systems evolved in recent years?
    \item[(2)] 
         What types of dependability threats are considered in the literature on UAV-based networks and computing systems?
    \item[(3)] 
         What techniques are adopted to improve the dependability of UAV-based networks and computing systems?
    \item[(4)] 
         What are the future research directions for dependable UAV-based networks and computing systems?
\end{itemize}

Based on these RQs, we developed a comprehensive search string encompassing three key conceptual components: UAV terminology, computing and networking domains, and dependability attributes. 
The final Boolean search string was designed as follows:

\begin{quote}
\texttt{("UAV" OR "UAS" OR "drone" OR "quadcopter" OR "aircraft") AND ("computing system" OR "network system" OR "edge computing" OR "fog computing" OR "cloud computing" OR "distributed system") AND ("dependability" OR "reliability" OR "availability" OR "safety" OR "fault-tolerance" OR "resiliency" OR "resilience" OR "survivability" OR "robustness" OR "performability")}
\end{quote}

We selected five major digital libraries as data sources: IEEE Xplore, ACM Digital Library, SpringerLink, ScienceDirect, and ISI Web of Science, and selected papers published between 2015 and 2024.

\subsection{Data Collection and Extraction}



Based on the selected databases, we first conducted an initial screening of articles using our structured search queries. 
This preliminary step aimed to retrieve a broad but relevant set of publications related to our survey scope. 
Subsequently, we applied the predefined inclusion and exclusion criteria to refine the pool of studies. 
The inclusion criteria focused on primary research that introduced novel methods, frameworks, or experimental insights addressing the dependability of UAV-based systems. 
We excluded studies primarily focused on cybersecurity, hardware-level reliability, or other secondary literature, such as review and survey papers, to ensure alignment with our scope.

The filtering process was conducted in two stages by multiple researchers: first, by reviewing titles and abstracts, and then by performing full-text screening. Any disagreements regarding article inclusion were resolved through consensus discussions to ensure consistency. 
For the final set of selected studies, we extracted detailed information relevant to our research questions, including addressed dependability threats, adopted metrics and methodologies, and application scenarios.

\subsection{Classification, Synthesis, and Reporting}

After data extraction, we systematically categorized the selected articles according to the views defined in our research questions. These aspects include system types (e.g., edge-connected UAVs), threat types (e.g., network disconnection), methodological approaches (e.g., modeling, optimization), and applications (e.g., monitoring, rescue). 
The general research trends on these keywords are  presented in Section~\ref{sec:research trends}.

Then, we synthesized the responses to research questions in Sections ~\ref{sec:dependability threats} and Section~\ref{sec:Dependability Techniques}, identifying dependability threats covered in the literature and dependability techniques to counteract these threats. Finally, based on the findings from comprehensive literature reviews, we discuss possible future research directions to answer RQ4 in  Section~\ref{sec:future research}.

\section{Research trends}
\label{sec:research trends}

To provide a comprehensive understanding of the dependability of UAV-based systems, we begin with an overview of the primary studies conducted in this field. 
Table \ref{tab:source_distribution} shows the number of studies identified by the search strings across selected databases, as well as the number of studies retained after applying the defined inclusion and exclusion criteria.
Initially, a total of 1,848 articles were retrieved from five databases.
After removing duplicates and applying the selection criteria, 458 articles were retained for full-text analysis and used to address the research questions.

\begin{table}[h]
\centering
\caption{Source-wise distribution of selected and accepted papers}
\begin{tabular}{|l|c|c|}
\hline
\textbf{Source} & \textbf{Selected} & \textbf{Accepted} \\
\hline
ACM Digital Library     & 19  & 7   \\ \hline
IEEE Digital Library    & 167 & 111 \\ \hline
ISI Web of Science      & 376 & 217 \\ \hline
Science@Direct          & 582 & 72  \\ \hline
Springer Link           & 704 & 51  \\
\hline
\end{tabular}
\label{tab:source_distribution}
\end{table}

\begin{figure}[h]
    \centering
    \includegraphics[width=0.8\linewidth]{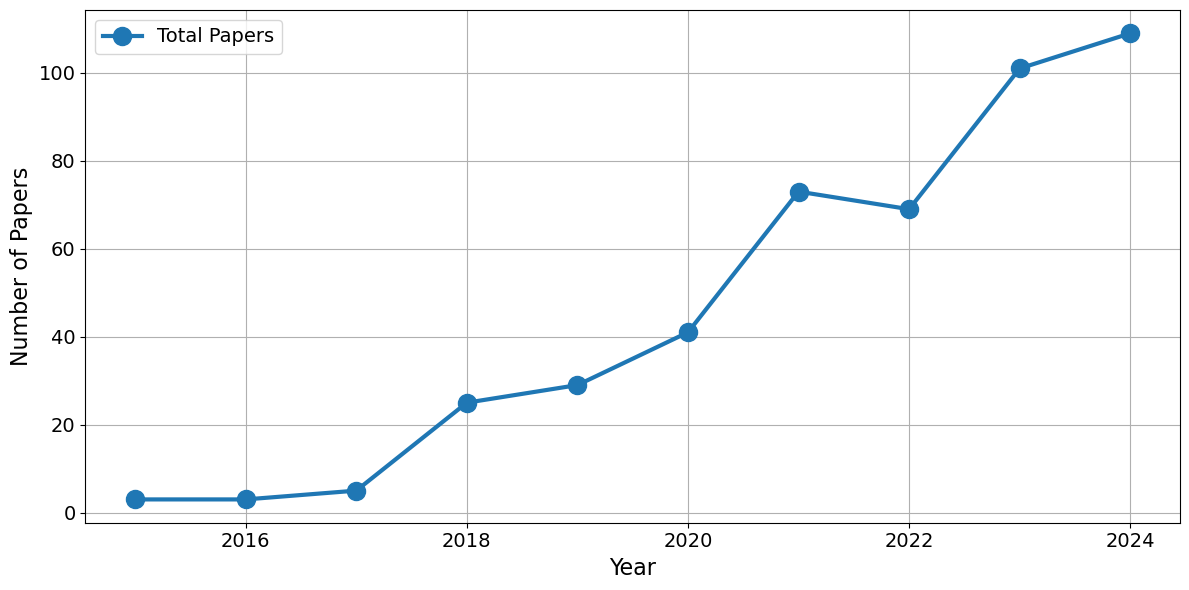}
    \caption{The number of articles per year}
    \label{fig:arti_stat}
\end{figure}

In particular, we survey the key papers published from 2015 to July 2024, which amount to 458 papers after filtering. These include  journal articles, conference proceedings, and some articles in open-source archives.
The analysis of publication trends over the years reveals a consistent and accelerating growth in research on dependability of UAV-based systems, as shown in Figure~\ref{fig:arti_stat}. 
The field began modestly, with only three publications in both 2015 and 2016. A noticeable increase occurred from 2018 onward, with 25 papers published that year. This growth continued steadily: 29 papers in 2019, 41 in 2020, and a sharp rise to 73 papers in 2021. These figures reflect the growing interest in UAV reliability, driven by technological advances and expanding application domains.
In recent years, the trend has intensified. In 2023 alone, 101 papers were published, followed by 109 publications in the first seven months of 2024. This numbers shows the increasing relevance of UAV dependability, reflecting their broader adoption in sectors such as logistics, surveillance, and communications.
This steady growth indicates a maturing and dynamic research area, marked by strong interdisciplinary contributions.


\subsection{Research trends}

To systematically analyze the research trends in the dependability of UAV-based systems, we closely examined the keyword frequencies and their temporal variations over the investigation period.

\subsubsection{Keyward frequency analysis}

\begin{figure}[htbp]
    \centering

    \begin{subfigure}[b]{0.33\textwidth}
        \centering
        \includegraphics[width=\textwidth]{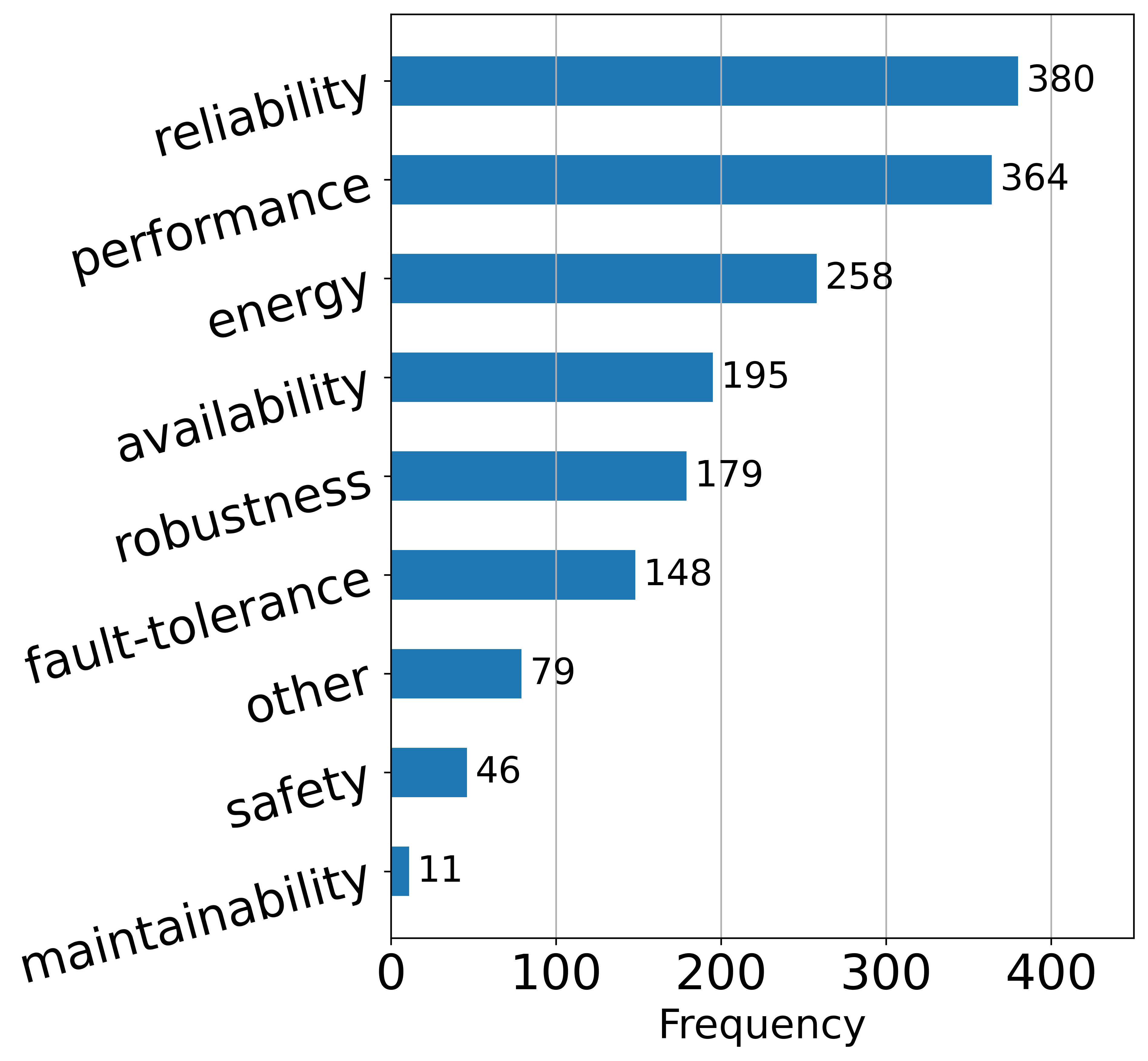}
        \caption{Dependability metrics}
        \label{fig:dm_a}
    \end{subfigure}
    \hfill
    \begin{subfigure}[b]{0.33\textwidth}
        \centering
        \includegraphics[width=\textwidth]{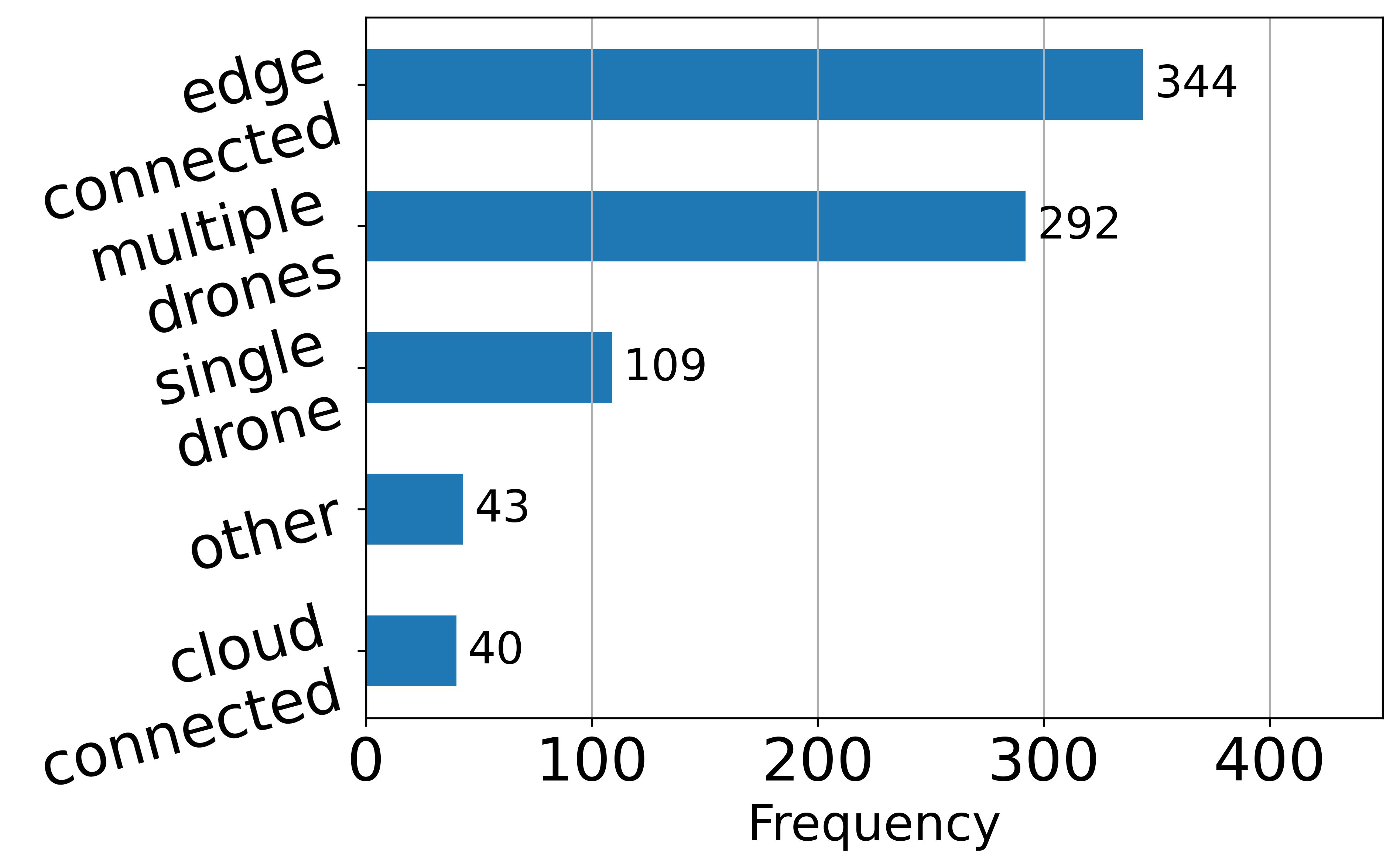}
        \caption{System type}
        \label{fig:dm_b}
    \end{subfigure}
    \hfill
    \begin{subfigure}[b]{0.33\textwidth}
        \centering
        \includegraphics[width=\textwidth]{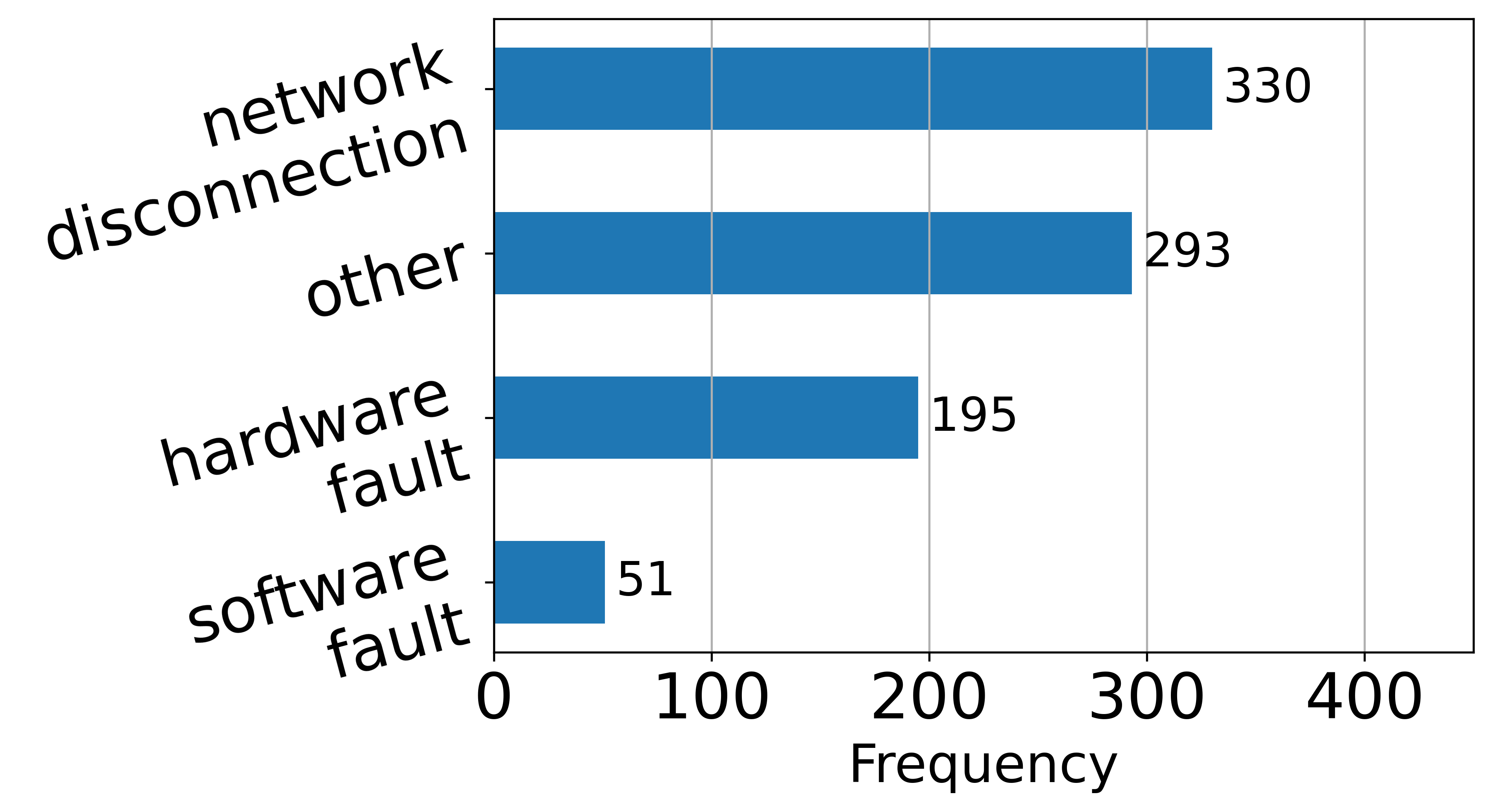}
        \caption{Dependability threats}
        \label{fig:dm_c}
    \end{subfigure}

    \vspace{0.6cm}

    \hspace{0.15\textwidth} 
    \begin{subfigure}[b]{0.33\textwidth}
        \centering
        \includegraphics[width=\textwidth]{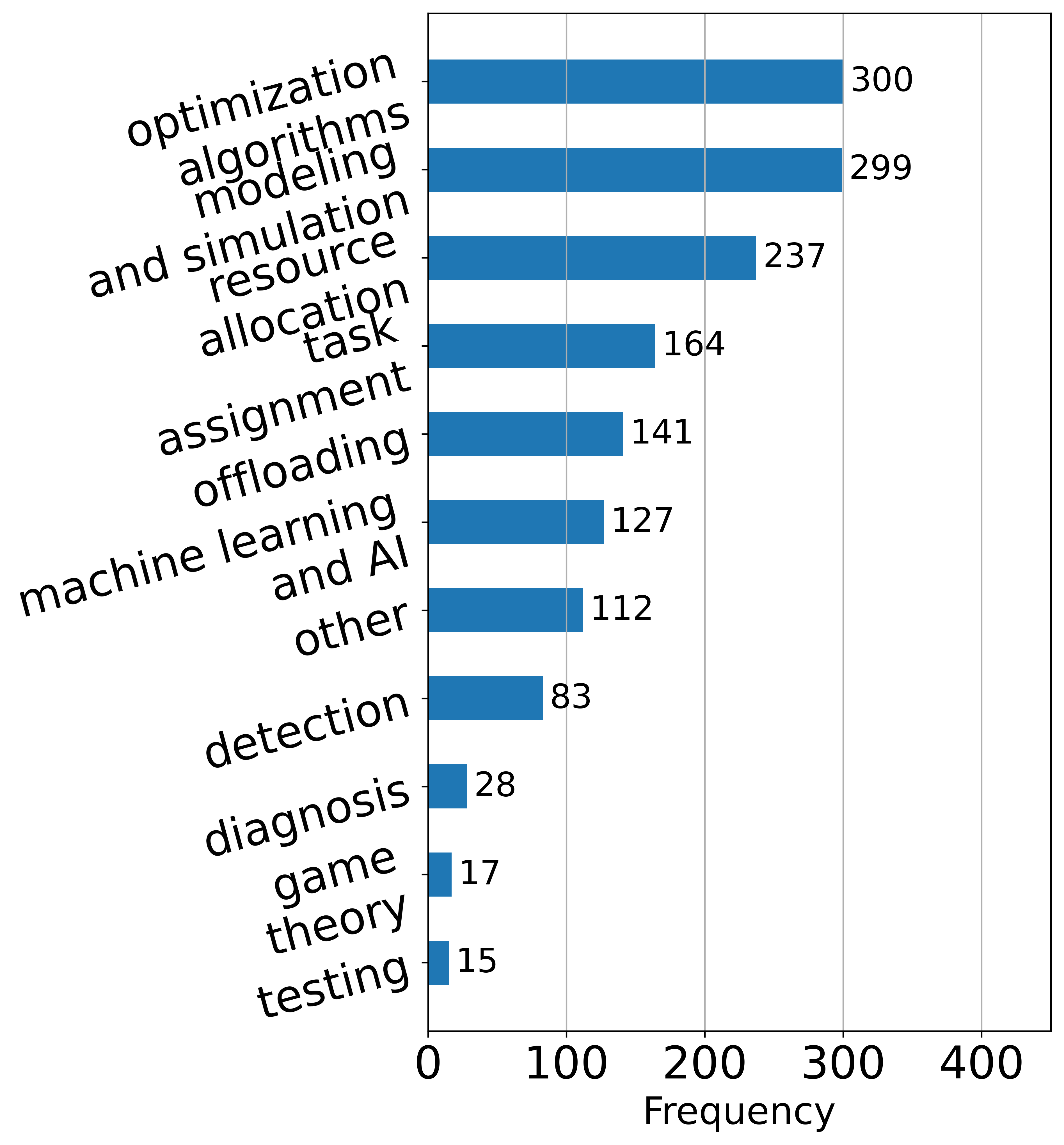}
        \caption{Methodological approaches}
        \label{fig:dm_d}
    \end{subfigure}
    \hspace{0.1\textwidth} 
    \begin{subfigure}[b]{0.33\textwidth}
        \centering
        \includegraphics[width=\textwidth]{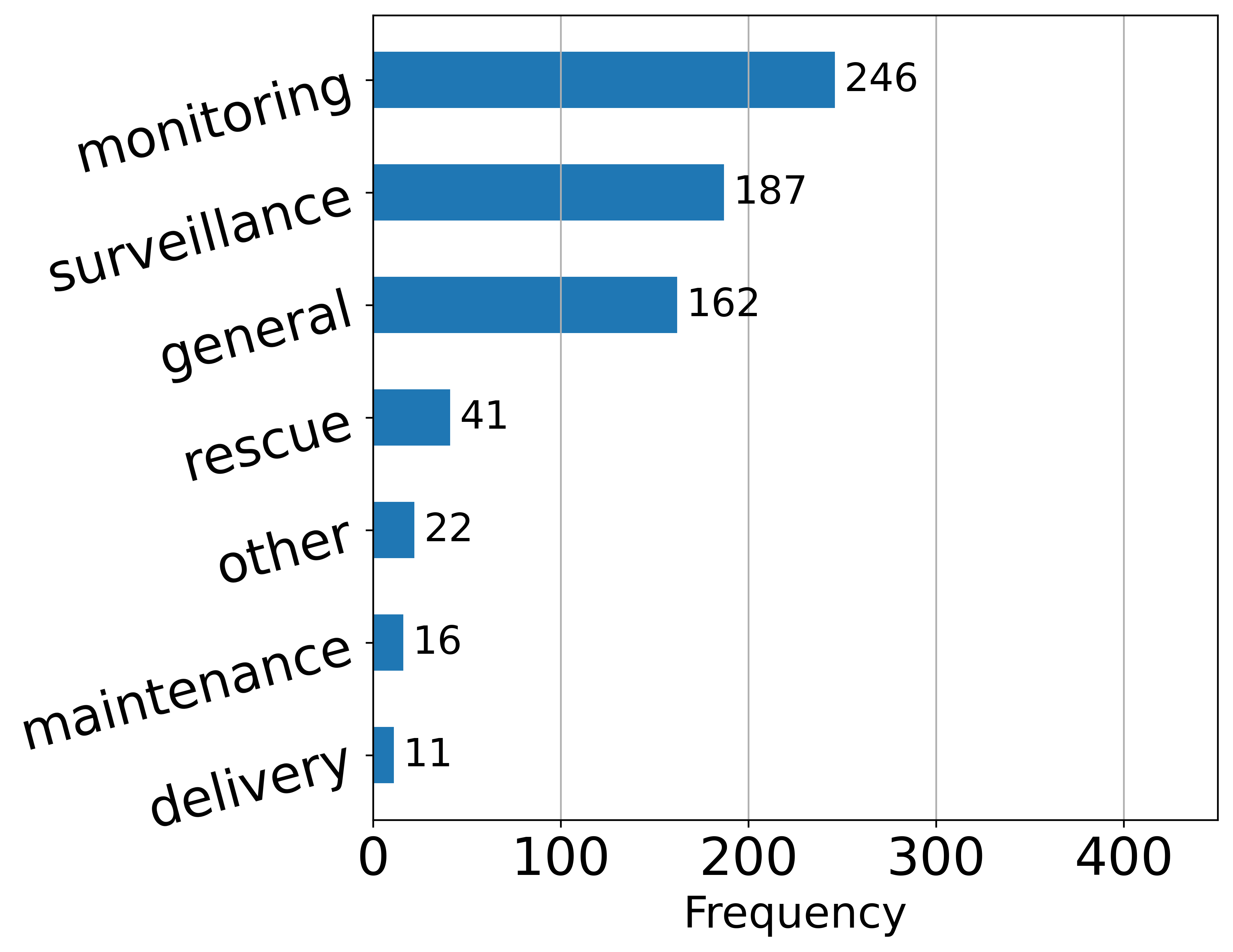}
        \caption{Applications}
        \label{fig:dm_e}
    \end{subfigure}

    \caption{Word frequencies across five major analysis aspects (a–e).}
    \label{fig:dm_all}
\end{figure}

First, we examined the total number of occurrences over the ten years for selected keywords in various subjects. The comparison of keyword frequencies highlights the hot areas discussed in the existing literature, as shown in Figure \ref{fig:dm_all}. In terms of dependability metrics shown in Figure \ref{fig:dm_a}, the most prominent emphasis is placed on reliability (380 mentions) and performance (364 mentions), exhibiting their foundational role in ensuring consistent UAV operations. Energy (258 mentions) and availability (195 mentions) also receive considerable attention, reflecting concerns around power constraints and operational readiness. Less frequently discussed metrics, such as robustness, fault tolerance, and especially safety and maintainability, highlight potential opportunities for future research in enhancing system resilience and post-deployment usability.

Regarding system types, as presented in Figure \ref{fig:dm_b}, edge-connected UAVs (344 mentions) and multi-drone systems (292 mentions) dominate the landscape, illustrating the growing importance of distributed, collaborative UAV networks. Single-drone systems, although less prevalent, remain relevant for specific applications, while cloud-connected architectures appear to be less emphasized, possibly due to latency or connectivity challenges. When analyzing dependability threats shown in Figure \ref{fig:dm_c}, the primary concern is network disconnection (330 mentions), reflecting the critical role of stable communication links in UAV-based systems. Hardware faults and software faults are also recognized as significant challenges, while the substantial frequency of the "other" category suggests a wide array of additional threats, such as environmental or operational factors.

In the realm of methodological approaches, as presented in Figure \ref{fig:dm_d}, optimization algorithms (300 mentions) and modeling \& simulation (299 mentions) are the dominant techniques employed to enhance UAV system performance and predictability. Other commonly used methods include resource allocation, task assignment, and offloading strategies. The notable application of machine learning, game theory, and detection and diagnosis techniques demonstrates an increasing trend toward intelligent, adaptive UAV systems. Finally, the application domain analysis in Figure \ref{fig:dm_e} reveals that monitoring (246 mentions) and surveillance (187 mentions) are the primary use cases for UAVs, emphasizing their critical role in security, environmental observation, and emergency response scenarios. Emerging applications such as rescue missions, maintenance, and delivery services appear less frequently, indicating potential areas for future expansion and innovation.

\subsubsection{Major research trends}

\begin{figure}[htbp]
    \centering
    \begin{subfigure}[b]{0.33\textwidth}
        \centering
        \includegraphics[width=\textwidth]{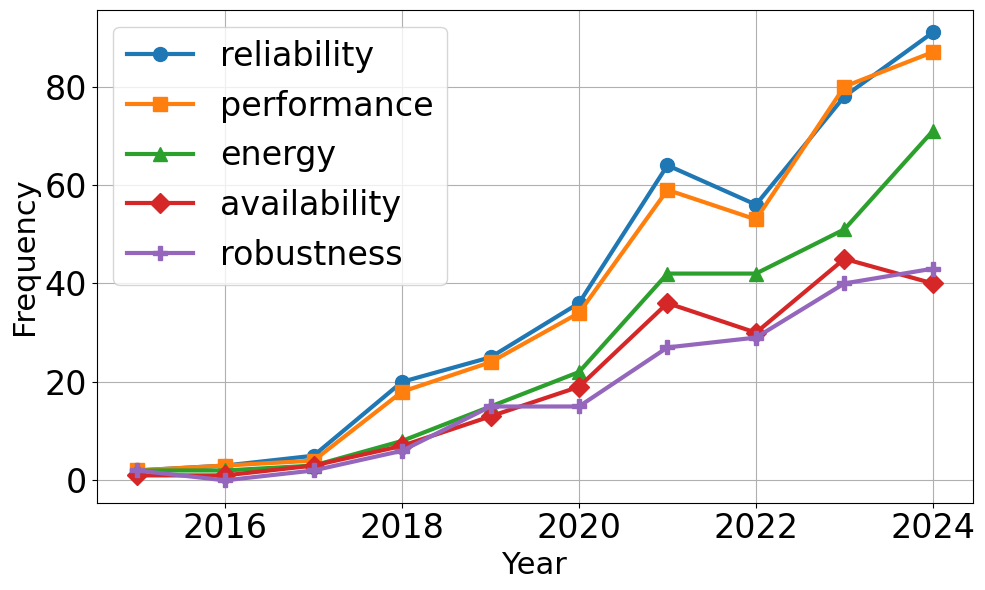}
        \caption{Dependability metrics}
        \label{fig:wf_year_dm}
    \end{subfigure}
    \hfill
    \begin{subfigure}[b]{0.33\textwidth}
        \centering
        \includegraphics[width=\textwidth]{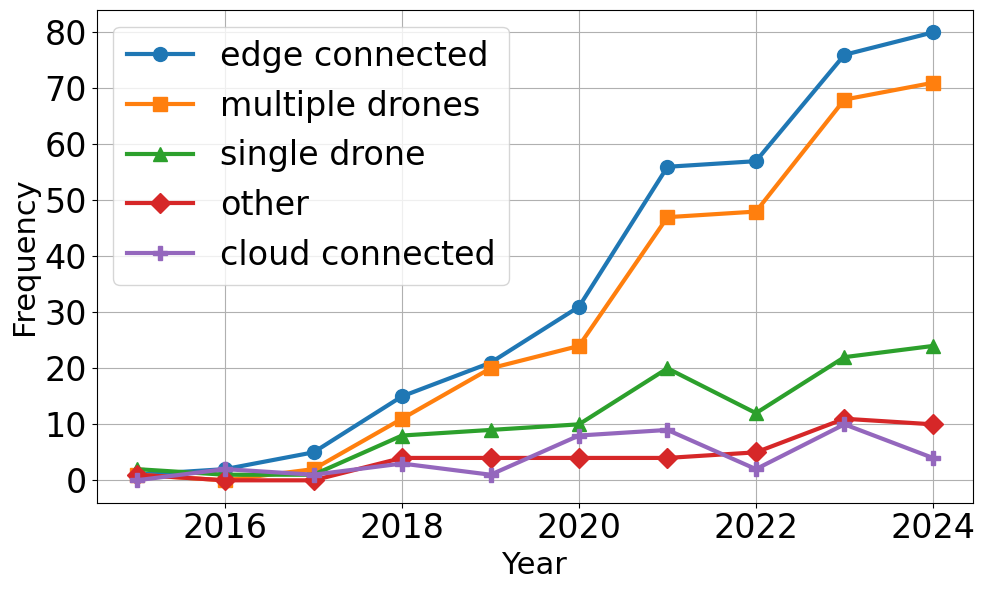}
        \caption{System type}
        \label{fig:wf_year_st}
    \end{subfigure}
    \hfill
    \begin{subfigure}[b]{0.33\textwidth}
        \centering
        \includegraphics[width=\textwidth]{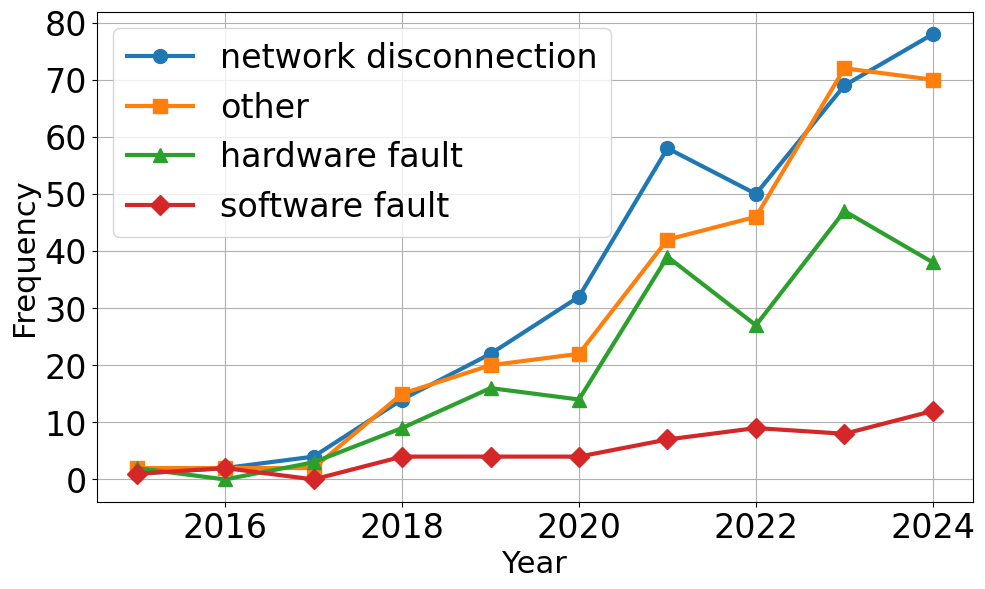}
        \caption{Dependability threats}
        \label{fig:wf_year_dt}
    \end{subfigure}

    \vspace{0.5cm}

    \begin{subfigure}[b]{0.33\textwidth}
        \centering
        \includegraphics[width=\textwidth]{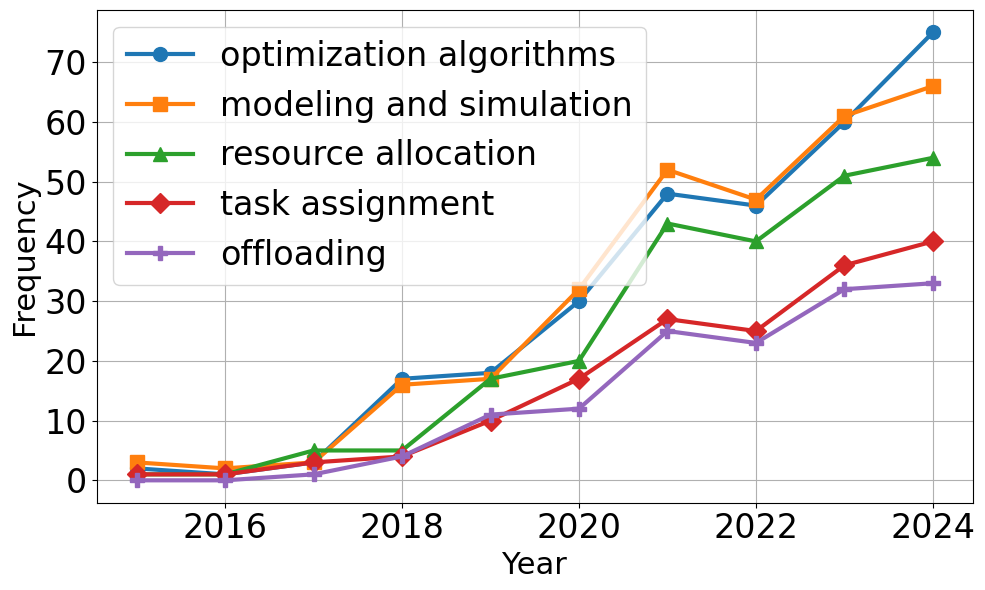}
        \caption{Dependability techniques}
        \label{fig:wf_year_ma}
    \end{subfigure}
    \hspace{0.1\textwidth} 
    \begin{subfigure}[b]{0.33\textwidth}
        \centering
        \includegraphics[width=\textwidth]{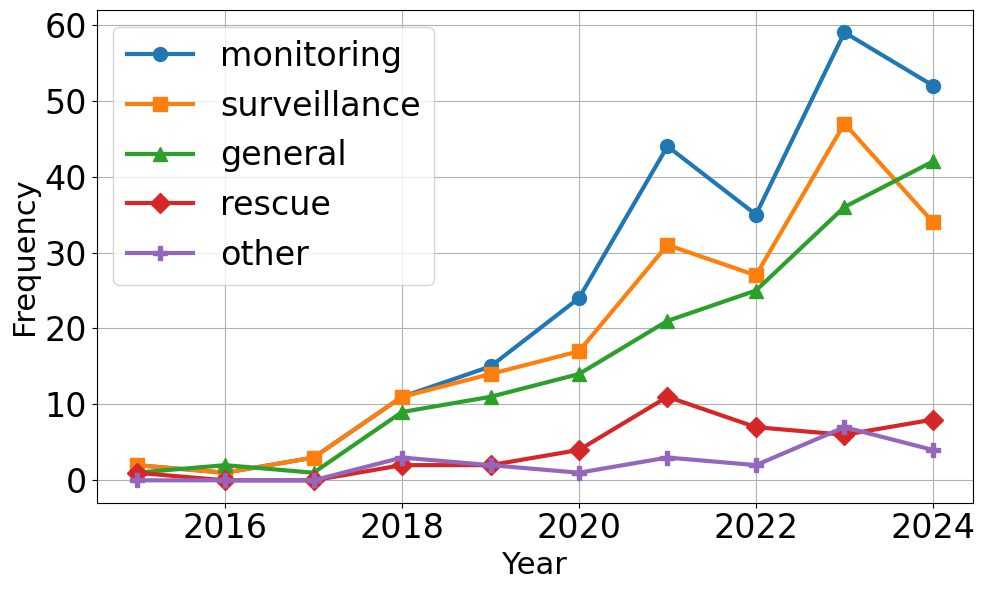}
        \caption{Applications}
        \label{fig:wf_year_a}
    \end{subfigure}

    \caption{Topic trends over the years across five categories: (a) metrics, (b) system types, (c) threats, (d) techniques, and (e) applications.}
    \label{fig:wf_year_all}
\end{figure}

To capture the evolving focus in different aspects, we conducted a temporal analysis of keyword frequencies over the years (see Figure \ref{fig:wf_year_all}).
By generating a year-wise pivot graphics, we systematically tracked how often each metric was mentioned in the literature from 2015 to 2024. This approach enables us to observe shifts in research priorities and emerging areas of interest over time.

The yearly distribution of \textbf{dependability metrics} reveals several notable trends, as shown in Figure \ref{fig:wf_year_dm}.
\textbf{Reliability} and \textbf{performance} consistently dominate the landscape, with mentions increasing from fewer than five in early years to over 90 and 87, respectively, by 2024. This indicates a sustained emphasis on ensuring the operational consistency and efficiency of UAV systems. \textbf{Energy}-related terms also demonstrate significant growth, reflecting increasing attention to power constraints and energy optimization, particularly as UAV missions become more complex. The frequency of \textbf{availability} and \textbf{robustness} similarly shows steady growth, highlighting a growing interest in system readiness and resilience to disruptions. Notably, mentions of \textbf{fault-tolerance} and \textbf{safety} also rise over the years, though at a comparatively moderate pace. \textbf{Maintainability} appears infrequently, suggesting it remains an underexplored aspect in the current research landscape. Overall, the results illustrate a maturing research focus, where foundational concerns, such as reliability and performance, remain important, while emerging challenges, like energy efficiency, fault handling, and safety, are receiving increased attention as UAV systems are deployed in more demanding and diverse environments.



The temporal distribution of \textbf{system types} reveals notable shifts in focus over the years, as shown in Figure \ref{fig:wf_year_st}. The most prominent trend is the steady increase in the attention to \textbf{edge-connected UAV systems}, with mentions rising from only one in 2015 to a peak of 80 by 2024. This indicates the growing importance of edge computing frameworks in supporting real-time processing, reducing latency, and enhancing decision-making capabilities, particularly as UAV applications become increasingly data-intensive.
Similarly, research on \textbf{multi-drone systems} has seen a significant rise, increasing from a few mentions in the early years to 71 mentions by 2024. The prominence of multi-drone studies reflects a clear emphasis on scalability, redundancy, and cooperative task execution, which are essential for enhancing the dependability of UAV systems in complex environments.
In contrast, \textbf{single-drone systems} maintain a relatively stable but lower frequency, suggesting their continued relevance for focused, cost-effective operations where simpler setups suffice.
Interestingly, \textbf{cloud-connected UAVs} show fluctuating yet consistently lower mentions throughout the years, peaking slightly at ten mentions in 2023. This may be attributed to the challenges associated with cloud dependence, such as latency and connectivity reliability, making edge-based solutions more attractive for mission-critical UAV tasks.
Lastly, mentions categorized under \textbf{other system types} remain modest, reflecting niche configurations or experimental system architectures tailored to specific use cases.
Overall, the results highlight a shift in research toward distributed, collaborative, and real-time capable UAV systems, aligning with broader trends observed in emerging UAV applications and technological advancements.



The yearly distribution of identified \textbf{dependability threats} in UAV systems highlights key vulnerabilities that have attracted sustained research attention, as shown in Figure \ref{fig:wf_year_dt}. The most frequently addressed threat is \textbf{network disconnection}, which shows a significant upward trend, increasing from only a few mentions in 2015-2016 to 78 mentions in 2024. This highlights the crucial importance of reliable communication links in UAV networks, particularly as these systems increasingly rely on real-time data exchange and edge or cloud connectivity. 
Similarly, concerns related to \textbf{hardware faults} have steadily increased, rising from occasional mentions in early years to a notable 47 mentions in 2023. This trend highlights the inherent susceptibility of UAV mechanical and electronic components to wear, environmental stress, and operational failures, particularly in demanding or prolonged missions. 
The \textbf{software fault} category, while receiving fewer mentions overall, exhibits a stable pattern across the years. The relatively lower frequency may suggest that software reliability issues, though important, are perceived as more controllable or well-understood compared to physical and network-related threats. 
The high and increasing counts in the \textbf{other threats} category—peaking at 72 mentions in 2023—indicate the diverse range of additional challenges faced by UAV systems, such as environmental factors, cyber-attacks, operational errors, and unpredictable mission conditions. This wide range underscores the complexity of ensuring dependability of UAV-based systems in real-world scenarios. 
In summary, the threat analysis reveals that network reliability and hardware robustness remain the primary concerns, while attention to software vulnerabilities and emerging threats continues to grow, suggesting an ongoing need for holistic and multi-layered dependability solutions.



The temporal analysis of \textbf{dependability techniques} reveals evolving preferences in the tools and techniques employed to enhance system performance and reliability, as shown in Figure \ref{fig:wf_year_ma}. 
\textbf{Modeling and Simulation} consistently dominates throughout the years, showing a strong and steady increase from three mentions in 2015 to 66 mentions in 2024. This highlights the importance of simulation-based approaches for testing UAV systems in diverse scenarios, allowing researchers to assess dependability without incurring real-world deployment risks. \textbf{Optimization Algorithms} similarly exhibit significant growth, with mentions rising from two in 2015 to 75 in 2024. This trend underscores the pivotal role of optimization techniques in enhancing UAV efficiency, resource management, and task scheduling. \textbf{Machine Learning and AI} methods display an exponential rise—from a minimal presence in early years to 32 mentions in 2024—indicating increasing integration of intelligent, adaptive systems to enhance UAV autonomy, fault prediction, and decision-making capabilities. Other noteworthy techniques include \textbf{offloading strategies}, which show a steady increase in mentions, emphasizing the need to manage computational loads effectively through edge or cloud resources. Mentions of \textbf{detection} and \textbf{diagnosis} techniques, though lower in frequency not shown in the picture, show gradual growth with single-digit, reflecting growing interest in real-time anomaly detection and system health monitoring. Interestingly, the application of \textbf{Game Theory}, while still emerging, with only five papers mentioned until 2023, has seen a noticeable increase in the last few years, suggesting its growing relevance in strategic decision-making, coordination, and resource negotiation within multi-UAV environments. Overall, the methodology trends indicate a progressive shift from traditional simulation-based validation towards the integration of AI-driven, optimization-centric, and distributed approaches, reflecting the complex operational requirements of modern UAV systems.



The temporal distribution of \textbf{UAV application }domains reveals clear priorities and emerging trends in the research landscape, as shown in Figure \ref{fig:wf_year_a}. Among the various application areas, \textbf{monitoring} and \textbf{surveillance} consistently dominate, with mentions increasing significantly over the years. \textbf{Monitoring}-related terms rose from just two mentions in 2015 to a peak of 59 mentions in 2023, while \textbf{surveillance} mentions grew from two in 2015 to 47 in 2023. This reflects the critical role of UAVs in environmental observation, security, and large-scale monitoring tasks, where dependability and operational readiness are paramount. The \textbf{general application} category also shows steady growth, increasing from minimal mentions in early years to 42 mentions by 2024. This indicates the versatility and expanding deployment of UAVs across diverse sectors, including those not confined to specific use cases. In contrast, mentions of \textbf{rescue missions} and \textbf{maintenance tasks}, though lower in absolute numbers, show gradual increases, highlighting the growing recognition of UAVs in safety-critical and infrastructure support roles. \textbf{Rescue missions}, for instance, reached 11 mentions by 2021, reflecting UAVs’ utility in emergency response scenarios. Interestingly, while \textbf{delivery applications} appear less frequently overall, there is a noticeable rise from 0 mentions in early years to occasional mentions in recent years. 
This suggests an emerging interest in UAV-based logistics, though it remains a relatively underexplored domain compared to surveillance and monitoring. The mentions categorized under \textbf{other applications} remain moderate, pointing to niche or evolving use cases tailored to specific operational needs.
In summary, the application analysis highlights the dominance of monitoring and surveillance, while also underscoring the growing opportunities in rescue, maintenance, and delivery domains that warrant further exploration.


\begin{tcolorbox}[colback=gray!10!white, colframe=black, title=\textbf{RQ1 – How have research trends on the dependability of UAV-based systems evolved in recent years?}]
Research trends on the dependability of UAV-based systems have shown significant evolution in recent years, reflecting the growing complexity and critical role of UAVs in diverse applications. The annual number of papers has surged from just three in 2015 to 109 in 2024, indicating a sharp increase in research interest, particularly since 2018, when the count rose from 25 to over 100 by 2023. 
Trends in dependability aspects reveal a consistent rise in attention to reliability, performance, and robustness, with the number of reliability papers increasing from two in 2015 to 91 in 2024, highlighting a shift toward comprehensive system performance. System types have evolved, with edge-connected and multiple-drone systems dominating since 2019, reflecting a move toward distributed and collaborative UAV networks. Dependability threats, such as network disconnections and hardware faults, have gained prominence, mirroring the real-world operational challenges of UAVs, paralleled by advancements like machine learning and optimization algorithms. Applications have expanded, with monitoring and surveillance leading by 2023, signaling a broadening scope of dependable UAV use.

\end{tcolorbox}

\section{Dependability threats}
\label{sec:dependability threats}

\begin{figure}
    \centering
    \includegraphics[width=1\linewidth]{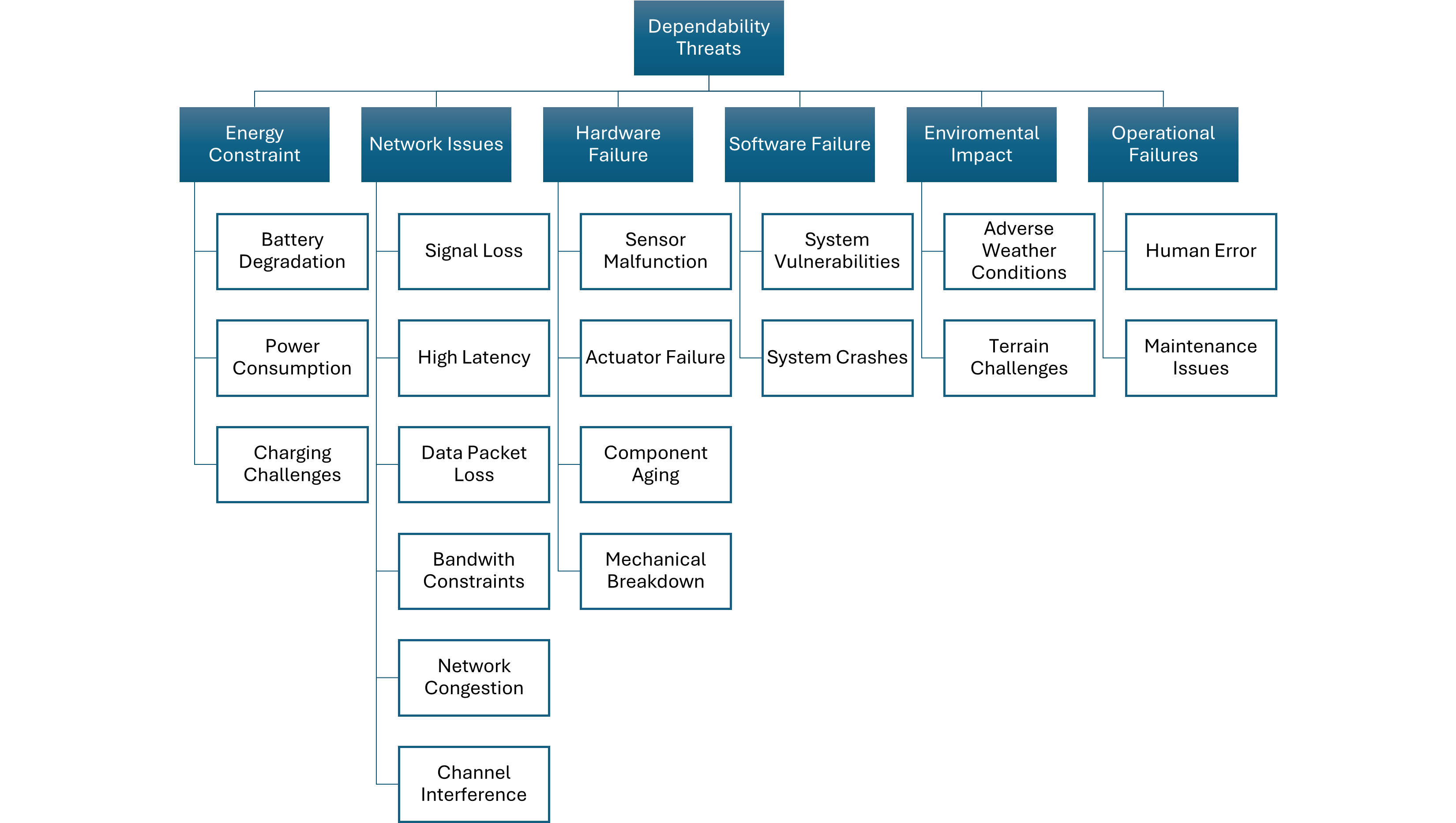}
    \caption{Dependability threats of UAV-based systems}
    \label{fig:enter-label}
\end{figure}

As UAVs take on increasingly mission-critical roles, ensuring their dependability becomes essential. This section examines core dependability challenges such as reliability, availability, integrity, and maintainability that directly affect UAV performance under both nominal and adverse conditions. Failures in these aspects can result in mission disruption, data loss, physical damage, or compromised safety, especially in dynamic or high-risk environments. Unlike traditional computing systems, UAVs operate in highly dynamic and often uncontrolled environments. They rely on the seamless integration of multiple subsystems such as sensors, actuators, power modules, communication networks, and embedded software to function properly. A failure in any one of these subsystems can propagate and result in system-wide consequences. Moreover, UAVs are subject to external conditions like weather, terrain complexity, and electromagnetic interference, which further exacerbate the risk landscape. To provide a structured understanding of these challenges, Figure  ~\ref{fig:enter-label} presents a comprehensive taxonomy of dependability threats in UAV systems. This taxonomy categorizes threats into seven major domains: (1) Energy Constraint, involving battery degradation, excessive power draw, and charging limitations; (2) Network Issues, such as signal loss, high latency, bandwidth limitations, congestion, and RF interference; (3) Hardware Failure, covering sensor malfunctions, actuator faults, aging components, and structural breakdowns; (4) Software Failure, including bugs, vulnerabilities, and system crashes; (5) Environmental Impact, focusing on adverse weather and terrain-related difficulties; and (6) Operational Failures, which stem from human error and maintenance oversights.

By systematically identifying and organizing these threats, the taxonomy in Figure~\ref{fig:enter-label} serves as a foundational reference for UAV designers, operators, and researchers. It supports more targeted risk assessments, improves fault tolerance planning, and guides the development of resilient UAV systems capable of maintaining dependable operation in real-world deployments. Table~\ref{tab:uav_dependability_categories} complements this taxonomy by categorizing the surveyed papers according to the identified dependability threat types in UAV-based systems, providing further evidence from the literature.

\begin{table}[htbp]
\centering
\caption{Papers categorization in dependability threats of UAV-based systems}
\label{tab:uav_dependability_categories}
\renewcommand{\arraystretch}{1.3}
\resizebox{\textwidth}{!}{%
\begin{tabular}{|p{3cm}|p{3.5cm}|p{7cm}|}
    \hline
    \textbf{Category} & \textbf{Subcategory} & \textbf{Related Studies} \\ \hline
    \multirow{3}{*}{Energy Constraint} 
    & Battery Degradation & \cite{Lu2022-eo}, \cite{Hou2020-zd}, \cite{Yang2022-jf}, \cite{Li2023-rk} \\ \cline{2-3}
    & Power Consumption & \cite{Falcao2024-zc}, \cite{Kumar2023-xz}, \cite{Alqudsi2023-hh} \\ \cline{2-3}
    & Charging Challenges & \cite{Liang2021-bd}, \cite{Huang2019-pl}, \cite{Hadi2023-le}, \cite{Kabashkin2023-rs} \\ \hline
    
    \multirow{6}{*}{Network Issues} 
    & Signal Loss & \cite{Zhang2022-yr}, \cite{Kramer2021-gf}, \cite{Li2020-yf}, \cite{Li2023-rk}     \\ \cline{2-3}
    & High Latency & \cite{Alam2019-yj}, \cite{Feng2023-qu}, \cite{Tang2023-oh}, \cite{Hou2020-wo}, \cite{Yu2021-ma}, \cite{Tang2022-ug}   \\ \cline{2-3}
    & Data Packet Loss & \cite{Mukherjee2023-ao}, \cite{Wang2019-vh}, \cite{Nwadiugwu2021-hd}, \cite{Jiang2019-ov}, \cite{Haber2019-mf}, \cite{Bai2023-cz}, \cite{Zhang2023-zv}, \cite{Chu2021-bd}, \cite{Mukherjee2020-am}    \\ \cline{2-3}
    & Bandwidth Constraints & \cite{Ali2022-gw}, \cite{Zhang2021-cz},  \cite{Wang2023-kb},  \cite{Tareq2018-bd} \\ \cline{2-3}
    & Network Congestion & \cite{Picano2024-vj}, \cite{Nwadiugwu2021-hd}, \cite{Picano2024-vj}, \cite{She2021-wf} \\ \cline{2-3}
    & Channel Interference & \cite{Shen2023-xp}, \cite{Wang2023-wb}, \cite{Sun2018-cv}, \cite{Zhang2024-gn},
    \cite{Li2020-yf}\\ \hline

    \multirow{4}{*}{Hardware Failure}
    & Sensor Malfunction & \cite{Saadaoui2023-fq}, \cite{Hou2020-wo}, \cite{Shen2023-xp}, \cite{Ju2024-sw}, \cite{Huang2024-sh}, \cite{Yuan2022-uk}, \cite{Chen2020-mg}, \cite{Kabashkin2023-rs} \\ \cline{2-3}
    & Actuator Failure & \cite{Amin2018-zb}, \cite{Zhou2018-kk}, \cite{Wang2019-vh}, \cite{Wang2021-qp}, \cite{Hirota2020-cd}, \cite{Hu2015-qd} \\ \cline{2-3}
    & Component Aging & 
    \cite{Shao2023-yd}, \cite{Aloqaily2021-xg}, \cite{Haber2019-mf}, \cite{Dogea2023-tt}, \cite{Ren2017-pi}, \cite{Guo2022-at}, \cite{Zheng2023-zp}, \cite{Wang2020-tk}, \cite{DAgati2024-yw}, \cite{9464045}, \cite{Xiong2019-wl}, \cite{Li2019-eb}, \cite{Navardi2023-xl}, \cite{Zhang2020-re}, \cite{Jiang2023-yf}, \cite{Zheng2021-ku}, \cite{Faraci2023-uz}, \cite{Liu2024-ug}, \cite{Seisa2022-sc}, \cite{Luo2021-bq}, \cite{Haber2019-mf}, \cite{Bao2023-rq}, \cite{Jiang2019-ov}, \cite{Falcao2023-vs}, \cite{Bai2023-of}, \cite{Jung2023-qh}, \cite{Ngoenriang2022-zv}, \cite{inproceedings}, \cite{Geng2024-pl}, \cite{Ning2024-vu}, \cite{De_Paula_Soares2023-ek}, \cite{Ren2024-hf}, \cite{He2023-at}, \cite{Gupta2022-ax}, \cite{Chen2024-rq}, \cite{Picano2024-vj}, \cite{Wang2021-ir},\cite{Li2023-rn}, \cite{inproceedings}, \cite{Khoramnejad2024-bj}, \cite{Zhou2024-qf}, \cite{Lei2023-lw}, \cite{Shao2023-yd}, \cite{Hou2020-zd}, \cite{Yu2023-gu}, \cite{Zhang2023-zv}, \cite{Tripathi2023-sh}, \cite{Lorincz2021-cp}, \cite{Pacheco2021-wy}, \cite{Patnayak2021-mx}, \cite{Pamuklu2023-kj}, \cite{Kaleem2019-dv}, \cite{Zeng2024-aq}, \cite{adil2024uav}, \cite{Hang2020-bv}, \cite{Wang2022-cj}, \cite{Li2024-pg}\\ \cline{2-3}
    & Mechanical Breakdown & \cite{Luo2021-bq}, \cite{Zhang2021-cz}, \cite{Russell2018-mv},  \cite{Shen2023-xp}, \cite{ Lv2024-qi}, \cite{Wang2020-jj}, \cite{Dai2024-du}, \cite{Choutri2022-rb}, \cite{Zhou2024-mn}, \cite{Lindqvist2020-ws}, \cite{Kabashkin2023-rs}, \cite{Lorincz2021-cp}, \cite{Behjat2019-rd} \\ \hline

    \multirow{2}{*}{Software Failure}
    & System Vulnerabilities & \cite{Pradhan2015-oa},
\cite{Wang2023-wb},
\cite{Ma2024-yx},
\cite{Vamvakas2019-ee},
\cite{Wang2023-wb},
\cite{Ren2017-pi},
\cite{Venkatagiri2018-qn},
\cite{Qu2022-zt},
\cite{Chang2024-bw},
\cite{Watanabe2022-as},
\cite{Zhong2021-uo},
\cite{Zuo2022-th},
\cite{Zhou2023-wn},
\cite{Chen2022-kh},
\cite{Yang2019-qm},
\cite{Hirota2020-cd}
    \\ \cline{2-3}
    & System Crashes & \cite{Ma2024-yx},
\cite{Amin2018-zb},
\cite{Hirota2020-cd},
\cite{Kramer2021-gf},
\cite{Liu2019-st},
\cite{Qu2022-zt},
\cite{Chang2024-bw},
\cite{Pan2024-ws},
\cite{Jennings2022-ds},
\cite{Shang2023-qv},
\cite{Taleb2023-al},
\cite{Zhang2021-em},
\cite{Grasso2022-ad},
\cite{Ritchie2020-fp},
\cite{Lin2021-an},
\cite{Mo2022-iq},
\cite{Gupta2021-eo},
\cite{Hong2022-fn},
\cite{Zhou2024-ov},
\cite{Zhong2021-uo},
\cite{Nguyen2023-pr},
\cite{Zhang2018-qz},
\cite{Kaymaz2024-kg},
\cite{Ahamad2024-rw},
\cite{Goel2021-ci},
\cite{Bai2023-cz}
 \\ \hline

    \multirow{2}{*}{Environmental Impact}
    & Adverse Weather Conditions & \cite{Wang2021-zc},
\cite{Wang2021-zc},
\cite{Gallego-Madrid2020-gl},
\cite{Hossain2017-db},
\cite{Pradhan2015-oa},
\cite{Haber2021-vz},
 \cite{Yang2022-jf},
\cite{Lee2022-sw},
\cite{Sankarasrinivasan2015-gu},
 \cite{Barnawi2021-xq},
 \cite{Mcilwaine2021-jc},
 \cite{Yang2022-jf},
 \cite{Fan2023-qs},
 \cite{Venkatagiri2018-qn},
  \cite{Kumar2023-xz}
 \\ \cline{2-3}
    & Terrain Challenges & \cite{Dai2018-eh},
\cite{Rokaha2019-za} ,
\cite{Paranjothi2019-ew},
\cite{Bakhtiari2020-bh},
\cite{Gao2022-bp},
\cite{Peng2023-qz},
\cite{Dai2018-eh},
\cite{Kim2024-dk},
\cite{Faraci2023-uz},
\cite{Mao2023-ag},
\cite{Shang2023-qv},
\cite{Lv2024-qi},
\cite{Wang2021-qp},
\cite{Hadi2023-le},
\cite{Zhou2024-mn}
 \\ \hline

    \multirow{2}{*}{Operational Failures}
    & Human Error & \cite{Xu2020-ip},
\cite{Hou2020-wo},
\cite{Yu2021-ma},
\cite{Liu2023-em},
\cite{Yuan2022-uk},
\cite{Falcao2023-vs},
\cite{Kim2024-dk},
\cite{Mao2023-ag},
\cite{Faraci2023-uz},
\cite{Thapliyal2024-us},
\cite{Zhang2024-gn},
\cite{Xu2020-ip},
\cite{Manzoor2021-fw},
\cite{Jeong2020-sk},
\cite{He2023-at},
\cite{Hui2024-vn},
\cite{Falcao2023-vs},
\cite{Li2024-hr},
\cite{Kaleem2019-dv},
\cite{Ozger2018-up},
\cite{Rao2024-is},
\cite{Ouyang2023-sa},
\cite{Ju2024-sw},
\cite{Tang2023-oh},
\cite{Amos2021-fi},
\cite{Shen2023-pk},
\cite{Wang2020-pr},
\cite{Vamvakas2019-ee},
\cite{Asim2022-so}
 \\ \cline{2-3}
    & Maintenance Issues & \cite{Guo2022-ro},
\cite{Jo2018-kn},
\cite{Wu2024-vs},
\cite{Lv2024-qi},
\cite{Yang2017-kv},
\cite{Ren2017-pi},
\cite{Zhang2020-re},
\cite{Pradhan2015-oa},
\cite{Lin2021-an},
\cite{Zhang2024-rb},
\cite{Patnayak2021-mx},
\cite{Shams2023-vn},
\cite{Huang2024-ew}
\\ \hline
\end{tabular}
}
\end{table}

\subsection{Energy Constraint}

Energy constraints critically affect UAV operations, which rely on limited onboard power. The main threats related to this category are presented below, highlighting challenges to ensure safe and efficient UAV functionality.

\begin{itemize}
\item \textbf{Battery Degradation:} Battery degradation refers to the gradual loss of battery capacity and performance due to repeated charging and discharging cycles. In UAVs, lithium-polymer (LiPo) batteries are the most commonly used due to their high energy density \cite{Lu2022-eo}. However, over time, these batteries undergo chemical aging, reducing the amount of charge they can hold \cite{Hou2020-zd}. This results in shorter flight durations \cite{Yang2022-jf}, inconsistent power output, and potential mid-flight failures \cite{Li2023-rk}. Factors that accelerate degradation include overcharging, deep discharges, and exposure to high temperatures. For instance, a UAV that initially offers 30 minutes of flight time may reduce to just 15 minutes after 100 charge cycles. Real-time monitoring and battery health management systems can help detect and mitigate degradation risks.

\item \textbf{Power Consumption:} Power consumption in UAVs is influenced by multiple subsystems, including propulsion, communication, payload sensors, and onboard computing. Propulsion, particularly in multi-rotor systems, is the largest power consumer, especially during aggressive maneuvers or hovering. High-definition video transmission and real-time processing for autonomous navigation also draw significant energy. Managing power consumption requires efficient hardware selection, software optimization, and flight path planning \cite{Falcao2024-zc}. For example, flying at a steady altitude with minimal changes in direction can conserve power, while rapid acceleration and deceleration will deplete the battery faster. In addition, virtualization overhead in MEC-enabled UAVs increases computational demands, leading to higher power consumption, which challenges the ability to maintain node availability and fulfill URLLC (ultra-reliable low-latency communication) requirements \cite{Falcao2024-zc}. Environmental factors, such as strong wind disturbances, also force UAVs to consume additional energy for stabilization and trajectory correction, further impacting overall efficiency \cite{Kumar2023-xz}. To address these issues, recent studies propose two energy-aware optimization schemes: one that minimizes flight cycles by optimizing sensor wake-up schedules, UAV trajectories, and time slots, and another that maximizes energy efficiency through the joint optimization of data transmission and propulsion power consumption \cite{Alqudsi2023-hh}.

\item \textbf{Charging Challenges:} Charging challenges refer to the difficulties in quickly and safely recharging UAV batteries in operational environments. In field conditions, access to stable power sources can be limited, and charging LiPo batteries requires specialized chargers with voltage and current regulation. Charging times are often long, reducing operational readiness \cite{Liang2021-bd}. Furthermore, improper charging can lead to overheating or battery swelling, posing safety hazards. Fast-charging solutions and battery swapping systems are being developed to reduce downtime, but they come with added weight and logistical complexity. For instance, in remote surveillance missions, UAV teams may need to carry generators or solar panels for recharging, adding to operational overhead. Recent studies have proposed optimizing UAV positions by considering recharging point accessibility and building avoidance to support energy management in real-world deployments \cite{Huang2019-pl}. A recent framework integrates intelligent reflecting surfaces (IRS), UAVs, and laser-based wireless charging to enhance energy harvesting efficiency, offering a new approach to reduce charging delays and support continuous UAV operation \cite{Hadi2023-le}. Additionally, service continuity challenges are tackled by modeling UAV replacement during charging periods to maintain network availability \cite{Kabashkin2023-rs}.
\end{itemize}

\subsection{Network Issues}

UAVs require stable and secure communication links. Below, we present the main threats in this category, which impact performance, safety, and mission reliability.

\begin{itemize}
\item \textbf{Signal Loss:} Signal loss is a critical issue for remotely piloted or semi-autonomous UAVs, resulting in the interruption of control commands and telemetry feedback. Causes include exceeding the communication range \cite{Zhang2022-yr}, line-of-sight obstructions (e.g., buildings, hills), antenna misalignment, hardware malfunctions, and network disconnection due to unstable communication links \cite{Kramer2021-gf}. When signal loss occurs, most UAVs trigger fail-safe behaviors such as returning to home or hovering in place . However, in GPS-denied environments or during complex missions, signal loss can result in mission failure or crashes. For example, a drone flying behind a tall building may lose the link with the operator and enter an unintended hover, exhausting its battery and eventually crashing. Recent studies have also highlighted that high Doppler shifts in UAV communication can cause signal fading and interference, degrading communication reliability. This issue is especially critical in high-mobility scenarios and may impact applications including but not limited to mobile edge computing (MEC) \cite{Li2020-yf}. Additionally, network disconnections between UAVs and control centers, often resulting from unstable links or system faults, further contribute to signal interruptions during flight missions \cite{Li2023-rk}.

\item \textbf{High Latency:} High latency refers to delays between command transmission and UAV response, which is especially problematic in real-time operations like inspection or FPV (First-Person View) control. Latency can be caused by long-range communication links (e.g., satellite), network congestion, or processing delays. In addition, limited onboard processing capacity often forces UAVs to offload computational tasks to the cloud, introducing further communication delays and increasing latency \cite{Alam2019-yj}. Conventional UAV image processing methods, which involve merging images and the use of ground control points (GCPs), can also introduce high latency and positioning errors if not managed carefully \cite{Feng2023-qu}. Moreover, the integration of UAVs and electric vehicles (EVs) in edge computing environments faces stringent latency challenges, especially under dynamic workloads and varying task demands from users \cite{Tang2023-oh}. Network disconnections between UAVs and user equipment have also been identified as a cause of increased latency in computational offloading, further impacting system reliability \cite{Hou2020-wo}. Additionally, intermittent connectivity and complex route optimization in UAV-enabled MEC systems contribute to high latency under dynamic network conditions \cite{Yu2021-ma, Tang2022-ug}. For instance, a 300 ms delay might not affect automated mapping missions but can make manual control erratic and dangerous. Reducing latency involves optimizing data routing, using low-latency protocols, and deploying edge-computing systems onboard the UAV.

\item \textbf{Data Packet Loss:} Packet loss occurs when one or more transmitted data packets fail to reach their destination, leading to gaps in communication. In UAVs, this can result in missing telemetry, sensor data, or delayed command execution. Causes include RF interference, poor signal quality, and buffer overflow in the UAV or ground station. In high-mobility networks such as FANETs (Flying Ad Hoc Networks), the likelihood of packet loss and intermittent connections increases due to dynamic topology changes and unstable links \cite{Mukherjee2023-ao}. Additionally, quadrotor UAV systems are vulnerable to network issues such as random delays and packet loss, which can destabilize real-time control and disrupt reliable wireless communication \cite{Wang2019-vh}. Enhancements at the MAC protocol level, such as prioritizing nodes based on proximity, have been proposed to reduce network delay and improve robustness against packet loss \cite{Nwadiugwu2021-hd}. Poor packet scheduling and resource allocation in multi-agent UAV systems can lead to higher packet loss, especially when network conditions change \cite{Jiang2019-ov}. Network disconnections between UAVs, ground stations, satellites, and IoT devices are also identified as major contributors to data loss and increased latency, impacting the reliability of computational offloading and communication services \cite{Haber2019-mf, Bai2023-cz, Zhang2023-zv, Chu2021-bd}. High packet loss can degrade mission accuracy. In IoDT applications, high message latency and packet loss can reduce quality of service (QoS), especially in dynamic environments \cite{Mukherjee2020-am}. Limited energy and memory in edge devices make these challenges worse. For example, a UAV sending image data back to a ground station may deliver incomplete or corrupted files, requiring retransmission and causing delays. Protocols with error correction and adaptive retransmission strategies help mitigate these issues.

\item \textbf{Bandwidth Constraints:} Bandwidth limitations affect the volume and speed of data transmission between the UAV and ground control. High-bandwidth applications like HD video streaming or large-scale sensor data collection require stable and fast connections. In multi-UAV operations or urban environments, limited bandwidth can lead to reduced data quality or interrupted connections \cite{Ali2022-gw, Wang2023-kb}. For example, a surveillance drone attempting to stream 4K footage over a congested 2.4GHz channel may suffer frame drops or lag. To address bandwidth challenges, frameworks like Fed-UAV implement federated learning on edge devices, reducing the amount of data transmitted to central servers and helping to minimize bandwidth usage \cite{Zhang2021-cz}. Another method to reduce bandwidth problems is to use a normalized link throughput strategy, which helps lower interference and improve data transmission \cite{Ali2022-gw}. SDN-based approaches are also used to handle network disconnections. Managing limited SBS resources and bandwidth helps ensure fast, reliable communication for AVs without relying too much on the cloud \cite{Tareq2018-bd}. 

\item \textbf{Network Congestion:} Network congestion happens when multiple UAVs or devices compete for the same communication resources, causing delays and data collisions. This is common in swarm operations, public events, or urban areas. Congestion leads to increased latency, packet loss, and sometimes total communication failure \cite{Picano2024-vj}. Lack of effective node prioritization in UAV networks can worsen congestion by increasing competition for limited communication channels \cite{Nwadiugwu2021-hd}. Heavy traffic and strict QoS demands on UAV and ground edge nodes can cause congestion, delay communications, and lead to missed deadlines for important tasks \cite{Picano2024-vj, She2021-wf}.  For instance, in a drone light show involving dozens of UAVs, congestion could cause synchronization errors, impacting performance and safety. Solutions involve dynamic channel allocation, time-division multiplexing, and efficient communication protocols that adapt to network load.

\item \textbf{Channel Interference:} Channel interference refers to the disruption of UAV communication by overlapping signals from other RF sources such as Wi-Fi routers, mobile phones, or other drones. Interference reduces signal clarity, increases error rates, and can cause unexpected disconnections \cite{Shen2023-xp}. Deliberate channel access attacks (CAA) and fast-varying channel conditions in dynamic environments, such as vehicular networks, both contribute to communication interference and instability in UAV networks \cite{Wang2023-wb, Sun2018-cv, Li2020-yf}. Channel uncertainty caused by dynamic air-ground communication in UAV networks also increases the risk of interference and affects the reliability of maintaining continuous connections, particularly for applications like the Metaverse \cite{Zhang2024-gn}. UAVs operating in crowded RF environments need robust modulation techniques and interference avoidance strategies. For example, a UAV inspecting infrastructure near a busy telecom tower may suffer from sporadic link quality due to interference. Frequency hopping and spectrum sensing technologies are commonly employed to reduce this risk.
\end{itemize}

\subsection{Hardware failure}

UAVs rely on reliable physical components to maintain stable and safe operation. Below, we present the main threats in this category, which impact flight control, system durability, and mission success

\begin{itemize}
\item \textbf{Sensor Malfunction:} Sensor malfunctions can occur in components such as GPS, IMU, altimeters, or cameras. These devices are critical for navigation, stabilization, and mission-specific tasks. Faulty or drifting sensors provide inaccurate data to the flight controller, leading to instability or incorrect responses  \cite{Saadaoui2023-fq}. For example, a GPS module reporting an incorrect position may cause the drone to veer off course, while an altimeter failure could result in improper altitude hold. Environmental factors such as magnetic interference or moisture can also contribute to sensor errors. UAV systems' sensor malfunctions can disrupt critical services such as mobile edge computing, vehicular communications, or fog computing operations, impacting overall system reliability \cite{Hou2020-wo, Shen2023-xp, Ju2024-sw, Huang2024-sh, Yuan2022-uk, Chen2020-mg}. Environmental conditions, such as bad weather or collisions, further increase the risk of sensor malfunctions, requiring fault management strategies like deploying additional UAVs to replace malfunctioning ones and maintain service continuity \cite{Kabashkin2023-rs}. To address these risks, system designs increasingly incorporate redundancy, regular calibration procedures, health monitoring, and fault-tolerant offloading strategies.

\item \textbf{Actuator Failure:} Actuators, such as servos and motor controllers, translate flight control commands into mechanical motion. Failure in these components can result in reduced maneuverability or total loss of control \cite{Amin2018-zb}. A jammed servo may lock a control surface on a fixed-wing UAV, while a faulty motor controller can shut down a rotor on a quadcopter. Main causes include overheating, mechanical wear, electrical faults, or broader hardware issues affecting actuators during flight operations  \cite{Zhou2018-kk, Wang2019-vh, Amin2018-zb}.  Some systems address actuator failures by enabling rapid re-planning in response to failures, allowing continued operation through dynamic adjustments \cite{Hirota2020-cd}. Low-cost UAVs are especially vulnerable due to limited actuator redundancy, making accurate fault detection and control switching essential; recent approaches use reliability frameworks with pivotal decomposition and new fault detection and isolation (FDI) metrics to handle actuator failures without relying on extra hardware \cite{Hu2015-qd}. Quadrotor UAV control systems are particularly vulnerable to actuator failures when combined with environmental disturbances and network issues like random delays and packet loss, which can threaten stability and disrupt real-time control \cite{Wang2021-qp}. Ensuring component quality, regular inspection, and incorporating fail-safes like motor redundancy in quadcopters can mitigate such risks.

\item \textbf{Component Aging:} With repeated usage, UAV components degrade due to mechanical fatigue \cite{Shao2023-yd}, corrosion, and thermal cycling. Bearings in motors, solder joints on PCBs, and electrical connectors are particularly vulnerable. Aging components can reduce efficiency, increase energy consumption, or suddenly fail mid-flight. 
In addition, long-term stress and environmental exposure can gradually weaken critical UAV hardware, increasing the risk of unexpected failures and service disruptions \cite{Aloqaily2021-xg, Haber2019-mf, Dogea2023-tt, Ren2017-pi, Guo2022-at, Zheng2023-zp, Wang2020-tk, DAgati2024-yw, 9464045, Xiong2019-wl}.
Such aging-related failures, combined with network disconnections between aircraft and ground stations, may result in increased latency or data loss, impacting the reliability of data transmission and service continuity. Efficient energy management is also critical, as aging components may draw more power, reducing operational time. Moreover, aging systems can be more prone to software faults, where bugs or vulnerabilities emerge over time, requiring robust maintenance and development practices to uphold dependability \cite{Ren2017-pi}.

\item \textbf{Mechanical Breakdown:} Structural components like frames, propellers, and landing gear are exposed to significant stress from takeoffs, landings, and potential collisions. Mechanical fatigue and wear over time can lead to cracks and fractures, particularly in components like landing gear and propeller mounts \cite{Luo2021-bq, Zhang2021-cz, Russell2018-mv}. Stress accumulation from repeated missions and vibrations contributes to deformation and weakening of airframes and connectors \cite{Shen2023-xp, Lv2024-qi, Wang2020-jj}. Mid-air collisions or impacts during landing are common triggers for abrupt structural failures, especially in multi-UAV or urban operations where maneuvering space is limited \cite{Dai2024-du, Choutri2022-rb, Zhou2024-mn, Lindqvist2020-ws}. Additionally, poor material quality or inadequate structural design increases the risk of breakage under operational loads \cite{Kabashkin2023-rs, Lorincz2021-cp, Behjat2019-rd}. A broken propeller mid-flight, for example, can result in immediate loss of control, particularly in systems lacking redundancy. Mechanical breakdowns are often preventable through visual inspections, structural stress testing, and the use of high-quality materials during design and assembly.
\end{itemize}

\subsection{Software failure}

UAVs depend on robust and secure software systems for navigation and control. Below, we present the main threats in this category, which affect system stability, security, and operational continuity.

\begin{itemize}
\item \textbf{System Vulnerabilities:} System vulnerabilities in UAVs refer to weaknesses in the software architecture, coding logic, or configuration settings that can be exploited by attackers or cause unintended behavior. These vulnerabilities may include unvalidated inputs, buffer overflows, and insecure memory handling, which expose the system to risks like memory corruption and code injection \cite{Pradhan2015-oa, Wang2023-wb, Ma2024-yx}. UAVs operating in sensitive environments such as surveillance or defense are especially at risk. For example, attackers may exploit unpatched firmware or default access credentials to hijack UAV functions or inject malicious commands \cite{Vamvakas2019-ee}. Some studies have also highlighted vulnerabilities in channel access mechanisms, where channel access attacks (CAA) can degrade the freshness and reliability of information by disrupting data transmission through protocol exploitation \cite{Wang2023-wb}. Additionally, system-level software faults, including bugs, logic errors, insecure configurations, integration issues within dynamic cyber-physical environments, or soft errors from environmental factors like radiation, can compromise functionality and security, disrupting service delivery and threatening mission continuity \cite{Ren2017-pi, Pradhan2015-oa, Ma2024-yx, Venkatagiri2018-qn, Qu2022-zt, Chang2024-bw}. Software aging is another critical concern, as prolonged operation may introduce subtle faults such as memory leaks or degraded performance, which can eventually lead to system failure or ineffective execution of key functions like task offloading \cite{Watanabe2022-as}. Furthermore, complex detection algorithms, such as those based on YOLOv5s, are vulnerable to configuration errors and mispredictions, which may result in high false alarm rates and reduced reliability of UAV-based detection systems \cite{Zhong2021-uo, Zuo2022-th, Zhou2023-wn}. Deep learning-based flight control systems may also encounter software faults, such as incorrect parameterization or model misbehavior in neural schedulers or DNNs, potentially leading to unsafe or inaccurate navigation decisions \cite{Chen2022-kh}. Monocular quadrotors also suffer from software limitations in depth estimation, where the absence of 3D information leads to unreliable obstacle avoidance, especially in unfamiliar or unstructured environments \cite{Yang2019-qm}. Some UAV systems implement fault-tolerant mechanisms that enable rapid re-planning in response to actuator or sensor faults and allow seamless path switching through precomputed recovery scenarios in the event of unexpected software errors \cite{Hirota2020-cd}. An attacker could also intercept unencrypted telemetry data and manipulate it to alter flight paths or collect confidential information. Regular software audits, secure coding practices, and real-time intrusion detection systems are essential to mitigate these risks.

\item \textbf{System Crashes:} UAV system crashes can occur due to software bugs, memory leaks, unexpected sensor inputs, or processing resources overload \cite{Ma2024-yx, Amin2018-zb, Hirota2020-cd}. These crashes may freeze the onboard computer, trigger automatic reboot, shut down critical flight functions or cause navigation-related faults \cite{Kramer2021-gf}. For example, a software loop that does not terminate might consume all of the system memory, leading to a crash during flight. If recovery systems are not well implemented, the UAV may fall or crash uncontrollably. Edge-enabled UAVs are also vulnerable to system crashes if compute offloading mechanisms fail or if the drone is left without sufficient onboard fallback, potentially halting autonomous functions during critical missions \cite{Liu2019-st}.  In multi-UAV edge orchestration frameworks, such as those used for collaborative video analytics, poorly optimized task scheduling or simultaneous high-load processing may overwhelm system resources, resulting in unresponsive drones or mid-mission system reboots \cite{Qu2022-zt}. Misconfigured control logic at edge terminals, especially when not detected by testing, can result in severe flight instabilities or control system crashes, as demonstrated in fuzzing-based evaluations of UAV safety risks \cite{Chang2024-bw}. Deep learning based detection modules like AIMED-Net, when deployed on resource-constrained edge platforms, may experience inference failures or overload under real-time constraints, potentially leading to unresponsive systems or control crashes during UAV operation. Comparative evaluations of onboard processors further highlight that under heavy computational loads, low-tier processors can become bottlenecks, increasing the risk of midmission system instability or crash \cite{Pan2024-ws, Jennings2022-ds}. Detection algorithms based on YOLOv5s can also suffer from configuration errors or inference instability, and in some cases, software crashes during execution, further increasing the risk of unresponsiveness in UAV detection systems\cite{Shang2023-qv, Taleb2023-al, Zhang2021-em, Grasso2022-ad, Ritchie2020-fp, Lin2021-an, Mo2022-iq, Gupta2021-eo, Hong2022-fn}. Other high-performance visual networks, such as VDTNet, may also introduce instability during edge-based detection and tracking tasks when deployed on resource-limited UAV platforms, potentially resulting in dropped detection, inference delays, or runtime crashes during intruder pursuit missions \cite{Zhou2024-ov}. System crashes can also stem from environmental hardware faults, energy depletion, unreliable communication links, or faults in AI-based detection and GPS/BDS-based attitude determination systems, which may cause incorrect orientation estimates and lead to UAV instability or crashes \cite{Zhong2021-uo, Nguyen2023-pr, Zhang2018-qz, Kaymaz2024-kg, Ahamad2024-rw}. Additionally, inaccurate estimation of stability and control derivatives from simulated flight data can result in poorly tuned controllers, increasing the risk of instability or system crash during real-world UAV operations \cite{Goel2021-ci}. A detailed fault tree of such failures in edge-based UAV orchestration has also been discussed in recent studies \cite{Bai2023-cz}. Reliability testing, hardware-in-the-loop simulations, and watchdog timers help manage the risk of software crashes.
\end{itemize}

\subsection{Environmental impact}

UAVs operate in complex and often harsh environments that can severely affect their performance and mission success. Below, we present the main environmental challenges that impact communication reliability, navigation accuracy, and overall system robustness.

\begin{itemize}
\item \textbf{Adverse Weather Conditions:} Environmental factors such as extremes of wind, rain, snow, fog, temperature, and dynamic air traffic can dramatically affect UAV performance and cause network disconnections that disrupt communication and data flow\cite{Wang2021-zc, Wang2021-zc, Gallego-Madrid2020-gl, Hossain2017-db, Pradhan2015-oa, Haber2021-vz}. For example, strong winds can destabilize flight, while rain can interfere with electronics and sensors that are not adequately protected \cite{Yang2022-jf, Lee2022-sw, Sankarasrinivasan2015-gu}. Cold weather reduces battery efficiency, limiting flight time, and icy conditions may add weight or damage propellers. Environmental factors also affect thermal sensor accuracy \cite{Barnawi2021-xq}. UAVs used for delivery or inspection must be ruggedized or equipped with weather-resilient sensors capable of adjusting flight behavior based on real-time input. Similarly, environmental factors such as sunlight glare, waves, and varying water color pose challenges for detection, while limited data availability further risks accuracy \cite{Mcilwaine2021-jc}. Wind gusts may also cause sudden attitude changes, while electromagnetic interference, particularly near industrial zones or high voltage lines, can corrupt communication signals or disrupt GNSS-based navigation \cite{Yang2022-jf}. Cloud cover and adverse conditions can cause data gaps or errors in mapping, reducing the reliability of timely agricultural monitoring \cite{Fan2023-qs}. In this context, weather forecasting and adaptive flight planning systems are essential to ensure mission reliability and operational continuity. Additionally, UAVs operating at high altitudes are vulnerable to soft errors caused by radiation exposure, which can induce silent data corruptions in memory or logic circuits, affecting onboard computation and processes such as video summarization. These transient faults may go undetected, leading to reliability degradation during critical missions. To mitigate these risks, systems often require redundancy mechanisms, error-correcting codes, or radiation-hardened hardware components \cite{Venkatagiri2018-qn}. Studies have shown that lightweight deep learning models such as MobileNet strike an effective balance between inference performance and power consumption, making them suitable for real-time trail detection in UAVs operating in adverse conditions \cite{Kumar2023-xz}.

\item \textbf{Terrain Challenges:} Terrain features such as mountains, forests, urban environments, or bodies of water can obstruct signals or limit UAV maneuverability, directly affecting operational reliability \cite{Dai2018-eh}. GPS signals may be reflected or blocked in canyons or dense urban areas, causing multipath errors and reduced navigational accuracy. Maintaining centimeter-level positioning accuracy becomes even more difficult under such conditions, particularly when using low-cost RTK-GNSS base stations \cite{Rokaha2019-za}. Tall buildings create radio control dead zones, while navigating tight spaces increases collision risks. In VANET-assisted UAV systems, urban obstacles such as buildings can disrupt radio waves, leading to frequent disconnections and failed message delivery \cite{Paranjothi2019-ew, Bakhtiari2020-bh}. UAVs face coverage challenges due to occlusions in geometrically complex areas like urban or mountainous terrains, which can degrade image quality \cite{Gao2022-bp, Peng2023-qz}. Limited onboard energy further necessitates efficient path planning to maximize mission effectiveness \cite{Dai2018-eh}. Additionally, detection systems deployed in remote or challenging terrains often face network disconnections, which can delay alerts and responses critical for low-altitude security operations \cite{Kim2024-dk, Faraci2023-uz}. Computational limits and channel obstacles hinder UAV task offloading, especially for dependent tasks modeled as directed acyclic graphs \cite{Mao2023-ag}. These challenges necessitate robust collision avoidance, terrain-following sensors like LiDAR, and autonomous systems to maintain reliable UAV operation in complex environments \cite{Shang2023-qv}. In addition to signal-related issues, hardware faults caused by environmental stress or mechanical wear in rugged terrains may lead to individual UAV failures during missions \cite{Lv2024-qi}. Furthermore, network disconnections in obstructed environments degrade coordination among UAVs, potentially disrupting formation maintenance and task execution \cite{Wang2021-qp}. Obstacles at ground level can also interrupt UAV-to-IoT communication, while limited battery capacity and hardware faults may disrupt edge computing services; laser-based power transfer has been proposed to stabilize such operations \cite{Hadi2023-le}. These challenges necessitate robust collision avoidance, terrain-following sensors like LiDAR, and autonomous systems to maintain reliable UAV operation in complex environments. Multi-UAV path planning also becomes difficult in uncertain, cluttered terrains where real-time obstacle avoidance and inter-UAV cooperation are critical for collision prevention and mission success \cite{Zhou2024-mn}.
\end{itemize}

\subsection{Operational failures}

UAV missions depend heavily on correct human operation and effective maintenance practices. Below, we present the main operational failure threats that affect system safety, reliability, and mission continuity.

\begin{itemize}
\item \textbf{Human Error:}

Human error is a leading cause of UAV accidents, often arising during mission planning, parameter input (e.g., altitude, geofencing), incorrect mode selection, or delayed responses in emergencies \cite{Xu2020-ip}. Typical scenarios include unintentionally disabling obstacle avoidance or launching missions with insufficient battery charge, which can result in system loss, property damage, or injury. Regular training and procedural checklists help mitigate such errors \cite{Hou2020-wo}. Recent studies highlight a range of human-involved or human-impacting operational vulnerabilities. In MEC-enabled UAV systems, operator-induced mistakes or insufficient monitoring can degrade system performance under tight latency and energy constraints \cite{Yu2021-ma, Liu2023-em, Yuan2022-uk, Falcao2023-vs}. Human misjudgment during UAV placement, route optimization, and offloading decisions may cause connectivity failures and task delays, especially in dynamic and uncertain network conditions \cite{Kim2024-dk, Mao2023-ag, Faraci2023-uz, Thapliyal2024-us}. Inaccurate assessment of battery status or communication strength also contributes to mission failure, particularly when UAVs serve dual roles (e.g., computing and relaying) \cite{Zhang2024-gn, Xu2020-ip, Manzoor2021-fw}. Scenarios involving limited pilot observation further complicate UAV status awareness, increasing the likelihood of unsafe maneuvers or late interventions \cite{Jeong2020-sk}. In high-demand or post-disaster environments, human error in managing energy and computing resources may result in service outages or imbalanced workloads \cite{He2023-at, Hui2024-vn, Falcao2023-vs, Li2024-hr, Kaleem2019-dv}. Moreover, inadequate responses to user-level interference (e.g., overloading UAV bandwidth) or security threats (e.g., eavesdropping) may stem from policy gaps or oversight limitations \cite{Ozger2018-up, Rao2024-is, Ouyang2023-sa, Ju2024-sw, Tang2023-oh, Amos2021-fi, Shen2023-pk}. Finally, system-level miscoordination due to delayed or misinformed human actions, such as poor handoff management or overutilized UAVs, leads to reduced QoE and compromised safety \cite{Wang2020-pr, Vamvakas2019-ee, Asim2022-so}. These scenarios underscore the importance of designing robust, human-aware systems, which include real-time feedback, operator training, and semi-autonomous error prevention mechanisms.

\item \textbf{Maintenance Issues:} Inadequate or skipped maintenance can lead to undetected issues such as loose connectors, worn-out rotors, degraded solder joints, or battery defects. Over time, these minor faults may escalate into critical in-flight failures, such as mid-air blade breakage. Ensuring dependability of UAV-based systems, therefore, demands regular inspections, maintenance documentation, and onboard diagnostic tools.
Maintenance concerns in drone swarms involve node-level failures, communication, structural, or mission-related issues that degrade collective reliability; complex network modeling is used to evaluate their impact \cite{Guo2022-ro}. UAV inspection systems face hardware wear and safety hazards, especially when image resolution or network stability limits detection reliability \cite{Jo2018-kn, Wu2024-vs}. Environmental factors such as weather and mechanical degradation introduce hardware faults that disrupt coordination and task execution, particularly in dynamic terrains \cite{Lv2024-qi}. Effective maintenance planning hinges on accurate failure forecasting. Time series models like Holt-Winters improve predictive accuracy, reducing downtime and safety risks \cite{Yang2017-kv}. Engine and component malfunctions in UAVs or aircraft, coupled with disconnections in diagnostic links, lead to increased latency and unreliable maintenance actions \cite{Ren2017-pi, Zhang2020-re}. Software bugs and vulnerabilities further necessitate continuous maintenance in cyber-physical and IIoT-based UAV systems \cite{Pradhan2015-oa, Lin2021-an}. System performance can also be affected by faults in RFID tags, wearable computing devices, or embedded processors, where network disconnections and energy limits hinder tracking and diagnostic accuracy \cite{Zhang2024-rb, Patnayak2021-mx}. In UAV-assisted robotic inspections, long-distance signal loss and energy constraints complicate real-time data processing and HD video transmission \cite{Shams2023-vn}. For offshore wind farm monitoring, limited battery life and long-range travel demand optimized maintenance routes and inspection scheduling \cite{Huang2024-ew}. 

\end{itemize}

\begin{tcolorbox}[colback=gray!10!white, colframe=black, title=\textbf{RQ2 – What types of dependability threats are considered in the literature on UAV-based networks and computing systems?}]

\begin{enumerate}
    \item  The literature identifies seven primary categories of dependability threats in UAV-based networks and computing systems: Energy Constraints, Network Issues, Hardware Failures, Software Failures, Environmental Impacts, Operational Failures, and Communication Breakdowns. Each category encompasses specific failure modes, such as battery degradation, signal interference, actuator malfunction, software crashes, and mission interruptions due to weather or human error.
    \item These threats arise from the integration of heterogeneous subsystems and the exposure of UAVs to dynamic, often uncontrollable environments. The review reveals that energy constraints (e.g., limited battery life) and network issues (e.g., latency, signal loss) are the most frequently addressed concerns, reflecting their critical influence on mission continuity and autonomous operation.
    
    \item The structured taxonomy developed in this chapter synthesizes the existing threat landscape and provides a foundation for targeted dependability analysis. It also highlights research gaps in underrepresented areas, such as long-term maintenance reliability and terrain-induced disruptions, offering direction for future investigations in dependable UAV system design.

\end{enumerate}

\end{tcolorbox}


\section{Dependability Techniques}
\label{sec:Dependability Techniques}
\begin{figure} [ht]
    \centering
    \includegraphics[width=1\linewidth]{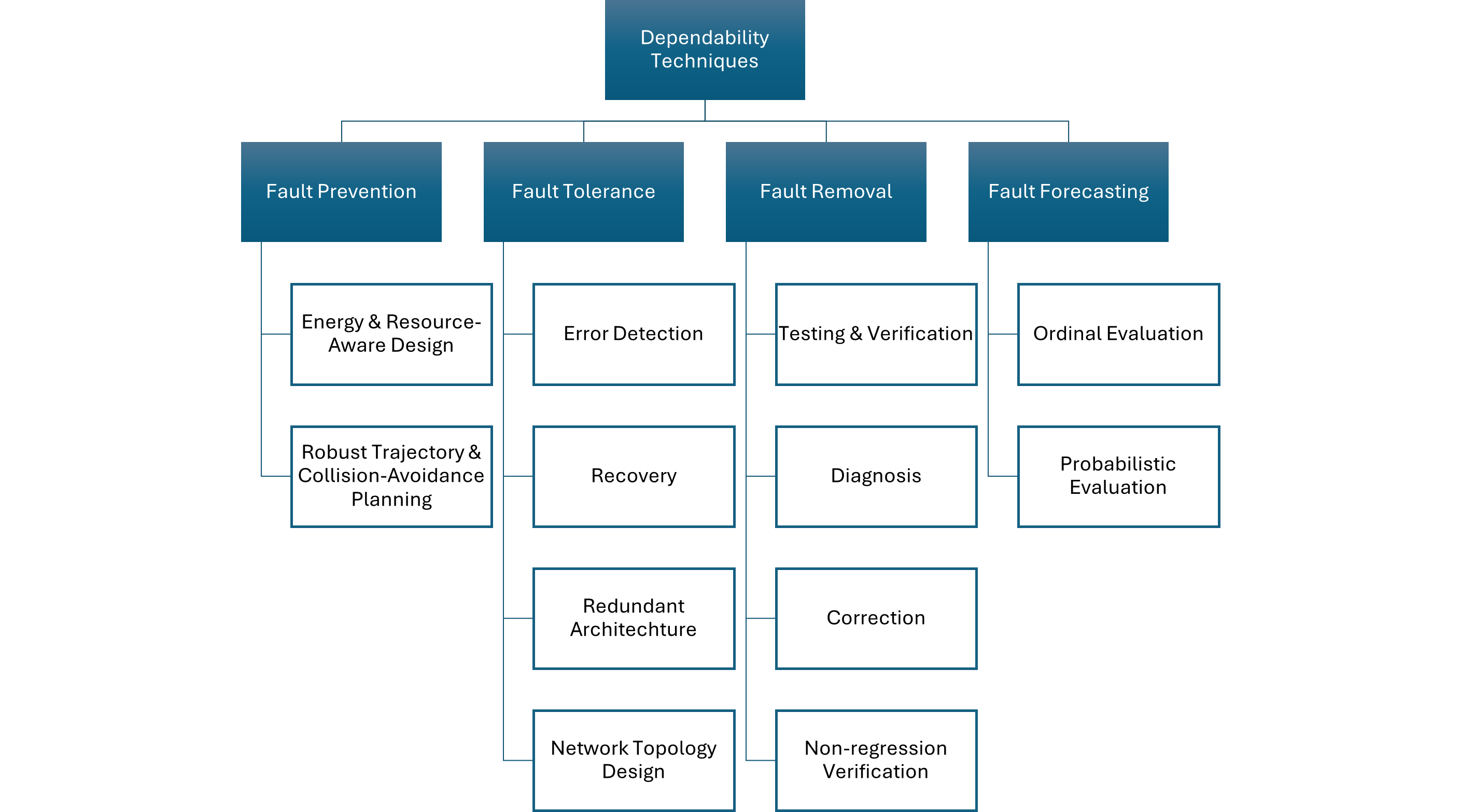}
    \caption{Dependability techniques for UAV-based systems}
    \label{fig:dependability-techniques-taxonomy}
\end{figure}

In the rapidly evolving field of UAVs, a wide array of methodologies and techniques has emerged to optimize performance, address challenges, and enhance the effectiveness of UAV systems in various applications. This section outlines the key dependability techniques that have been extensively explored in the literature, categorizing them into four principal categories as shown in Figure \ref{fig:dependability-techniques-taxonomy}. Each category represents a unique set of strategies and methods that contribute to the improvement of UAV capabilities. Within each of these categories, various subcategories and specific techniques have been developed and refined to address particular challenges, such as minimizing task completion time, ensuring reliable data transmission, or optimizing UAV energy consumption. These techniques not only focus on improving specific aspects of UAV performance but also enable more adaptable systems capable of responding to dynamic environments and real-time decision-making. As we examine each category in detail, we also explore how these techniques relate to each other and highlight trends that have emerged in recent research.

To systematically analyze the dependability techniques in UAV-enabled computing systems, the selected papers were categorized into four major groups derived from the foundational classification proposed by Avizienis et al. (2004): \textbf{Fault Prevention}, \textbf{Fault Tolerance}, \textbf{Fault Removal}, and \textbf{Fault Forecasting}. Each group consists of several subcategories, which reflect the specific methods or strategies employed in the reviewed papers.

\subsection{Fault Prevention}
Fault prevention includes techniques that aim to eliminate or reduce the likelihood of faults before they occur. These techniques are typically applied during the design and development stages of a system to ensure that faults are less likely to occur or be triggered during runtime. In the context of UAV-enabled environments, fault prevention is particularly important due to limited computing and energy resources, as well as the safety-critical nature of many UAV missions. The reviewed literature reveals a growing emphasis on prevention strategies that improve reliability and safety by addressing architectural, operational, and environmental vulnerabilities. Table \ref{tab:fault_prevention} summarizes the subcategories identified in fault prevention and the corresponding related studies.

\begin{table}[htbp]
\centering
\caption{Categorization of Fault Prevention Techniques in UAV-Based Systems}
\renewcommand{\arraystretch}{1.3}
\begin{tabular}{|c|p{4cm}|p{7.5cm}|}
\hline
\textbf{Technique} & \textbf{Subcategory} & \textbf{Related Studies} \\
\hline
\multirow{2}{*}{\textbf{Fault Prevention}} 
& Energy \& Resource-Aware Design & \cite{Ramos-Ramos2024-hq}, \cite{Zhou2022-zr}, \cite{Wei2022-oq}, \cite{Jeong2020-sk}, \cite{Zhang2024-kh}, \cite{Fatima2024-tx}, \cite{Maatouk_Letaifa_Rachedi_2024}, \cite{Xu2020-hx}, \cite{Suganya2024-hd}, \cite{Peng2024-tj}, \cite{Palossi2019-zv}, \cite{Rahul2021-bu}, \cite{Wang2022-ik}, \cite{Asim2022-so}, \cite{Zhang2023-js}, \cite{Alioua2018-mv}, \cite{Zhang2025-xl}, \cite{Zhejiang}, \cite{Selim2021-lo}, \cite{Xue2023-fq}, \cite{Tang2024-vt}, \cite{Wang2020-oy}, \cite{Laroui2021-fp}, \cite{Lin2021-an}, \cite{Zhang2021-ph}, \cite{Raj2023-ig}, \cite{Shao2023-yd}, \cite{Seisa2022-sc}, \cite{Hossain2017-db}, \cite{Lu2024-bz}, \cite{Grasso2019-ur}, \cite{Dou2024-qq}, \cite{Ma2022-to}, \cite{Qureshi2024-al}, \cite{Grasso2022-gx}, \cite{Gupta2021-eo}, \cite{Wang2019-mi}, \cite{Chen2020-ca}, \cite{Awada2023-xi}, \cite{Navardi2023-xl}, \cite{Martinez-Sanchez2023-uu}, \cite{Velez2024-gf}, \cite{Faraci2023-uz}, \cite{Zeng2024-aq}, \cite{Shi2023-fi}, \cite{Hong2022-fn}, \cite{Grasso2022-ad}, \cite{Baktayan2022-dj}, \cite{Hang2020-bv}, \cite{Tsung2022-ur}, \cite{Cheng2023-xn}, \cite{Asheralieva2020-er}, \cite{Geng2024-pl}, \cite{Shams2023-vn}, \cite{Gao2019-st}, \cite{Akbari2024-zs}, \cite{Li2024-pg}, \cite{Hadjkouider2023-po}, \cite{Zheng2024-zk}, \cite{Karam2024-eu}, \cite{Xiang2024-va}, \cite{Venkatagiri2018-qn}, \cite{Liu2019-st}, \cite{Callegaro2021-dg}, \cite{Khan2022-pj}, \cite{Li2020-wh}, \cite{Zhao2022-nr}, \cite{Xu2024-wl}, \cite{Asheralieva2019-uo}, \cite{Yu2022-lp}, \cite{Huang2024-em}, \cite{Apostolopoulos2023-sw}, \cite{Jung2023-qh}, \cite{Zhang2018-ss}, \cite{Wang2023-kb}, \cite{Haber2021-vz}, \cite{Pandey2021-tn}, \cite{Wang2024-cf}, \cite{Dai2024-du}, \cite{Alam2019-yj}, \cite{Subburaj2023-gq}, \cite{Fatima2024-tx}, \cite{Yang2020-bx}, \cite{Gu2022-gl}, \cite{Liu2023-em}, \cite{Falcao2024-zc}, \cite{Wang2020-tk}, \cite{Hou2020-zd}, \cite{Lu2022-eo}, \cite{Yu2021-ma}, \cite{Kim2021-uy}, \cite{Manzoor2021-fw}, \cite{Zhang2024-gn}, \cite{Thapliyal2024-us}, \cite{Awada2022-cm}, \cite{Li2020-zs}, \cite{Li2024-hr}, \cite{Tang2024-kd}, \cite{Luo2023-oe}, \cite{Lin2022-lq}, \cite{Ghosh2023-qq}, \cite{Islam2019-jn}, \cite{Ye2020-sg}, \cite{Yu2022-zp}, \cite{Ju2024-sw}, \cite{Ouyang2023-sa}, \cite{Rao2024-is}, \cite{Awada2023-xi}, \cite{Zhang2024-gb}, \cite{Rokaha2019-za}, \cite{Hao2024-wh}, \cite{Tang2022-gi}, \cite{Zhang2023-jy}, \cite{Pan2022-ka}, \cite{Xu2022-la}, \cite{Liu2023-ny}, \cite{Haber2019-mf}, \cite{Wang2024-zc}, \cite{Liang2021-bd}, \cite{Tang2024-sl}, \cite{Zeng2023-ee}, \cite{Wang2021-ir}, \cite{Hou2020-wo}, \cite{Chu2021-bd}, \cite{Tang2022-ug}, \cite{Ju2024-sw}, \cite{Chen2020-nc}, \cite{Yuan2022-uk}, \cite{Falcao2023-vs}, \cite{Liu2024-ak}, \cite{Hossain2022-yj}, \cite{Chen2020-mg}, \cite{Xue2023-ml}, \cite{Hadi2023-le}, \cite{Gao2020-xc}, \cite{Kumar2023-xz}, \cite{Pandey2021-if}, \cite{Jeong2020-sk}, \cite{He2023-at}, \cite{Han2021-bl}, \cite{Dai2022-mg}, \cite{Zhang2024-xq}, \cite{Wang2021-wl}, \cite{Zhang2022-yr}, \cite{Huang2024-ew}, \cite{Zhang2022-yw}, \cite{Wei2021-gh}, \cite{Qu2022-zt}, \cite{Chai2023-ke}, \cite{Wang2018-yk}, \cite{Zhang2023-vd}, \cite{Zhang2024-pq}, \cite{Asim2023-zp}, \cite{Ouyang2024-ik}, \cite{Seong2024-pt}, \cite{Yu2023-nk}, \cite{Awada2022-cm}, \cite{Geng2024-pl}, \cite{Awada2021-vl}, \cite{Zeng2024-aq}, \cite{Zhang2023-zv}, \cite{Hou2020-wo}, \cite{Khoramnejad2024-bj}, \cite{Chen2024-rq}, \cite{Wang2020-tk}, \cite{Zheng2023-zp}, \cite{Chu2021-bd}, \cite{Li2024-pg}, \cite{Qureshi2024-al}, \cite{Hou2020-zd}, \cite{Wang2024-cf}, \cite{Tian2024-ku}, \cite{Zhang2018-ss}, \cite{Bezziane2024-wu}, \cite{Chen2021-ft}, \cite{Basharat2023-fs}, \cite{Chen2022-kh}, \cite{Zou2022-no}, \cite{DAgati2024-yw}, \cite{Zhao2024-ke}, \cite{Su2022-ds}, \cite{Thapliyal2024-us}, \cite{Gao2019-st}, \cite{Ma2022-to}, \cite{Akbari2024-zs}, \cite{Xiong2019-wl}, \cite{Al-habob2024-zk}, \cite{Sun2020-fa}, \cite{Navardi2023-xl}, \cite{Faraci2023-uz}, \cite{Seisa2022-sc}, \cite{Luo2021-bq}, \cite{Haber2019-mf}, \cite{Bao2023-rq}, \cite{Bai2023-of}, \cite{Jung2023-qh}, \cite{Ngoenriang2022-zv}, \cite{Kong2022-uh}, \cite{Ning2024-vu}, \cite{He2023-at}, \cite{Gupta2022-ax}, \cite{Picano2024-vj}, \cite{Wang2021-ir}, \cite{9464045}, \cite{Kong2022-pl}, \cite{Shao2023-yd}, \cite{Pamuklu2023-kj}, \cite{Kaleem2019-dv}, \cite{Lin2021-an}, \cite{Garge2018-yk}, \cite{Hang2020-bv}, \cite{Tendolkar2021-qi}, \cite{Hui2024-vn}, \cite{Xue2023-qx}, \cite{Gao2024-ms}, \cite{Manzini2023-ir}, \cite{Liu2024-ak}, \cite{Ju2023-ed}, \cite{Pang2024-bk} \\
\cline{2-3}
& Robust Trajectory \& Collision-Avoidance Planning & \cite{Lee2022-zd}, \cite{Krishnan2021-dn}, \cite{Alqudsi2023-hh}, \cite{Wang2022-ik}, \cite{Hu2024-dl}, \cite{Asim2022-so}, \cite{Zhang2023-js}, \cite{Xu2024-vu}, \cite{Singh2023-qs}, \cite{Souli2023-ea}, \cite{Zhejiang}, \cite{Li2024-fo}, \cite{Kaymaz2024-kg}, \cite{Zeng2024-ig}, \cite{Hirota2020-cd}, \cite{Gao2021-bp}, \cite{Chen2022-kh}, \cite{Zheng2024-zk}, \cite{Liu2019-st}, \cite{Huang2024-em}, \cite{Wang2023-kb}, \cite{Choutri2022-rb}, \cite{Dai2024-du}, \cite{Subburaj2023-gq}, \cite{Liu2023-em}, \cite{Wang2020-tk}, \cite{Hou2020-zd}, \cite{Lu2022-eo}, \cite{Yu2021-ma}, \cite{Luo2021-bq}, \cite{Tripathi2023-sh}, \cite{Tang2024-kd}, \cite{Lee2022-sw}, \cite{Luo2023-oe}, \cite{Dai2018-eh}, \cite{Ye2020-sg}, \cite{Behjat2019-rd}, \cite{Koerkamp2019-ib}, \cite{She2021-wf}, \cite{Huang2019-pl}, \cite{Ju2024-sw}, \cite{Awada2023-xi}, \cite{Li2018-gv}, \cite{Deng2022-nm}, \cite{Zhou2024-mn}, \cite{Baccarelli2024-qs}, \cite{Li2024-ou}, \cite{Kim2024-zm}, \cite{Shen2023-xp}, \cite{Huang2024-sh}, \cite{Xue2023-ml}, \cite{Peng2022-dw}, \cite{Lv2024-qi}, \cite{Qayyum2021-qx}, \cite{Yang2022-jf}, \cite{Lorincz2021-cp}, \cite{Wang2021-wl}, \cite{Yang2019-qm} \cite{Lindqvist2020-ws}, \cite{Ye2022-td}, \cite{Guo2022-at}, \cite{Bekkouche2018-lt}, \cite{Tripathi2023-sh}, \cite{De_Paula_Soares2023-ek}, \cite{Zhan2024-dw}, \cite{10.1145/3551662.3561200}, \cite{Li2024-hr}, \cite{Gao2021-bp}, \cite{Hirota2020-cd}, \cite{Wu2023-qy}, \cite{Zheng2024-zk}, \cite{Jiang2019-ov}, \cite{Ngoenriang2022-zv}, \cite{Kong2022-uh}, \cite{Ning2024-vu}, \cite{Wang2024-px}, \cite{Marcos2021-ea} \\
\hline
\end{tabular}
\label{tab:fault_prevention}
\end{table}

\subsubsection{Energy \& Resource-Aware Design:}
Energy \& Resource-Aware Design focuses on strategies that proactively manage power consumption and computational workload to ensure stable UAV operations, particularly under limited energy and processing capacity. This approach is essential in UAV-based computing systems, where energy depletion or overload can lead to mission failure or system shutdown. This technique primarily addresses energy exhaustion, system overload, and thermal-related hardware degradation, which are critical dependability threats in UAV deployments. By carefully allocating resources, adjusting processing rates, or offloading tasks to nearby nodes or edge servers, UAVs can operate more sustainably and with greater resilience.

For instance, energy-aware offloading frameworks are proposed to reduce UAV energy consumption by dynamically assigning tasks to ground servers or other UAVs based on current battery levels and computational demands \cite{Ramos-Ramos2024-hq, Zhou2022-zr, Fatima2024-tx, Wang2022-ik, Zhejiang, Suganya2024-hd, Peng2024-tj, Zhang2025-xl, Wang2020-oy, Seisa2022-sc, Chen2020-ca, Faraci2023-uz, Asheralieva2020-er, Li2024-pg, Liu2019-st, Callegaro2021-dg, Jung2023-qh, Haber2021-vz, Yang2020-bx, Awada2022-cm}. Additionally, many works integrate power control and bandwidth allocation schemes to adaptively regulate UAV transmissions and processing loads, thereby lowering overall energy usage while maintaining service quality \cite{Zhang2024-kh, Xu2020-hx, Selim2021-lo, Tang2024-vt, Hossain2017-db, Hang2020-bv, Cheng2023-xn, Gao2019-st, Xiang2024-va, Zhao2022-nr, Zhang2018-ss, Wang2023-kb, Liu2023-em, Kim2021-uy}. These strategies directly contribute to enhancing system availability and reliability by preventing unexpected shutdowns and maintaining functional operation under varying load conditions.

\subsubsection{Robust Trajectory \& Collision-Avoidance Planning:}
Robust Trajectory \& Collision-Avoidance Planning refers to the proactive design of UAV flight paths that minimize the risk of collision, unsafe proximity, or environmental hazards. This technique is crucial for ensuring operational safety and fault prevention in autonomous UAV missions, especially in dynamic or congested airspaces. This technique directly mitigating threats related to mechanical breakdowns, terrain challenges, adverse weather conditions, and human error. Studies in this area typically employ trajectory optimization algorithms and collision-avoidance schemes to maintain navigational integrity and system continuity.

Multiple studies have proposed robust trajectory planning and avoidance frameworks that balance safety with performance. For example, model predictive control, velocity obstacles, and tabu search methods are commonly used to generate safe flight paths in constrained environments \cite{Lee2022-zd, Wang2022-ik, Xu2024-vu, Bekkouche2018-lt, Wu2023-qy}. Reinforcement learning and DRL-based approaches further enhance adaptivity, allowing UAVs to dynamically adjust paths in response to moving obstacles or real-time conditions \cite{Hu2024-dl, Behjat2019-rd, Gao2021-bp, Kong2022-uh, Ning2024-vu}. Some works also incorporate probabilistic models or risk-aware strategies to handle uncertainties such as wind gusts, localization errors, or multi-agent interactions \cite{Zeng2024-ig, Liu2019-st, Luo2021-bq, Zhan2024-dw}. 

In high-density or mission-critical scenarios like disaster recovery or automated surveillance, multi-UAV coordination algorithms are introduced to prevent in-air collisions while minimizing delays or energy usage \cite{Singh2023-qs, Lee2022-sw, Li2018-gv, Jiang2019-ov}. These techniques contribute directly to dependability metrics such as safety and reliability, enabling UAV systems to maintain continuous operation even in volatile or adversarial environments.

\subsection{Fault tolerance}
Fault tolerance includes techniques that enable UAV-based computing systems to continue functioning correctly even when faults occur during operation. These techniques are essential in dynamic and resource-constrained UAV environments where failures in hardware, software, or communication components may be inevitable. By incorporating mechanisms such as error detection and recovery, fault tolerance ensures system availability, reliability, and robustness in the presence of unexpected disruptions. The reviewed literature highlights various approaches that allow UAV systems to autonomously detect and recover from faults to maintain dependable mission execution. Table \ref{tab:fault_tolerance} summarizes the subcategories identified in fault tolerance and the corresponding related studies.

\begin{table}[htbp]
\centering
\caption{Categorization of Fault Tolerance Techniques in UAV-Based Systems}
\renewcommand{\arraystretch}{1.3}
\begin{tabular}{|c|l|p{7cm}|}
\hline
\textbf{Technique} & \textbf{Subcategory} & \textbf{Related Studies} \\
\hline
\multirow{4}{*}{\textbf{Fault Tolerance}} 
& Error Detection & \cite{Ahamad2023-ni}, \cite{Cheng2024-fu}, \cite{Vemulapalli2021-vf}, \cite{Gokulakrishnan2023-qw}, \cite{Ma2024-yx}, \cite{Saadaoui2023-fq}, \cite{Kostoglotov2024-lp}, \cite{Zhang2018-qz}, \cite{Zhou2024-ov}, \cite{Amin2018-zb}, \cite{article-jeong}, \cite{Wang2021-zc}, \cite{Mo2022-iq}, \cite{Chang2024-bw}, \cite{Alam2019-yj}, \cite{Wu2024-vs}, \cite{Balamuralidhar2021-un}, \cite{Behjat2019-rd}, \cite{Bakhtiari2020-bh}, \cite{Rokaha2019-za}, \cite{Nwadiugwu2021-hd}, \cite{Zhong2021-uo}, \cite{Jeong2016-cz}, \cite{Shang2023-qv}, \cite{Pan2024-ws}, \cite{Li2024-fp}, \cite{Zhou2023-wn}, \cite{Zhang2024-rb}, \cite{Zheng2021-ku}, \cite{Yuwei2020-ex} \\
\cline{2-3}
& Recovery & \cite{Farrukh2023-lp}, \cite{Luo2022-jt}, \cite{Bobronnikov2022-yn}, \cite{Alkinani2021-az}, \cite{Zuo2022-th}, \cite{Wang2021-qp}, \cite{Zhou2018-kk}, \cite{Callegaro2021-dg}, \cite{Choutri2022-rb}, \cite{Wang2024-cf}, \cite{Hou2020-zd}, \cite{Luo2021-bq}, \cite{Machida2021-jc}, \cite{Kim2021-uy}, \cite{Li2024-hr}, \cite{Machida2023-hq}, \cite{Yu2022-zp}, \cite{Ju2024-sw}, \cite{Awada2023-xi}, \cite{Zhang2024-gb}, \cite{Machida2023-hq}, \cite{Petrenko2023-ly}, \cite{Pan2022-ka}, \cite{Watanabe2022-as}, \cite{Wang2019-vh}, \cite{Li2023-mz}, \cite{Damigos2023-fa}, \cite{Machida2021-jc} \\
\cline{2-3}
& Redundant Architecture & \cite{Ma2024-yx}, \cite{Luo2020-ye}, \cite{Farrukh2023-lp}, \cite{La_Salandra2024-dr}, \cite{Ren2017-pi}, \cite{Kilbourne2021-xv}, \cite{Nguyen2023-pr}, \cite{Patnayak2021-mx}, \cite{Awada2021-vl}, \cite{Zhang2021-em}, \cite{Gallego-Madrid2020-gl}, \cite{Kaleem2019-dv}, \cite{Bian2023-ev}, \cite{Dogea2023-tt}, \cite{Karam2024-eu}, \cite{Aloqaily2021-xg}, \cite{Liu2019-st}, \cite{Callegaro2021-dg}, \cite{Khan2022-pj}, \cite{Zhao2022-nr}, \cite{Jung2023-qh}, \cite{Sharma2019-ah}, \cite{Pacheco2021-wy}, \cite{Hou2020-zd}, \cite{Awada2022-cm}, \cite{Ali2022-gw}, \cite{Kim2024-ix}, \cite{Petrenko2023-ly}, \cite{Sun2018-cv}, \cite{El_Sayed2023-mf}, \cite{Sara2016-zc}, \cite{Chen2024-up}, \cite{Barnawi2021-xq}, \cite{Wang2020-pr}, \cite{Javanmardi2024-tj}, \cite{Panwar2023-zc}, \cite{Shen2023-pk}, \cite{Naouri2024-uw}, \cite{Tang2023-oh}, \cite{Pustokhina2021-ae}, \cite{Taleb2023-al}, \cite{Patnayak2021-mx}, \cite{Lorincz2021-cp}, \cite{Karam2024-eu}, \cite{Lei2023-lw}, \cite{Li2023-rn}, \cite{Haber2021-vz}, \cite{Li2023-rk}, \cite{Pradhan2015-oa} \\
\cline{2-3}
& Network Topology Design & \cite{Xu2021-hb}, \cite{Guruprasad2024-aj}, \cite{Zhang2023-zb}, \cite{Vamvakas2019-ee}, \cite{Wang2024-xo}, \cite{Chen2022-qh}, \cite{Fujii2017-zp}, \cite{Shao2023-yd}, \cite{Martinez-Sanchez2023-uu}, \cite{Wang2021-ow}, \cite{Dao2021-gw}, \cite{Ozger2018-up}, \cite{A_Alissa2023-ev}, \cite{Tan2020-hc}, \cite{Su2021-vt}, \cite{Khan2021-jq}, \cite{Wang2022-cj}, \cite{Zhou2024-qf}, \cite{Kong2022-me}, \cite{Chen2024-hm}, \cite{Aloqaily2021-xg}, \cite{Khan2022-pj}, \cite{Li2020-wh}, \cite{Haber2021-vz}, \cite{Sharma2019-ah}, \cite{Gu2022-gl}, \cite{Thapliyal2024-us}, \cite{Li2020-yf}, \cite{Zhang2021-cz}, \cite{Jung2023-gj}, \cite{Mukherjee2020-am}, \cite{Paranjothi2019-ew}, \cite{Peng2023-qz}, \cite{Nwadiugwu2021-hd}, \cite{Mukherjee2023-ao}, \cite{Mao2023-ag}, \cite{Zhang2023-zb}, \cite{Yue2018-by}, \cite{Yadav2023-lw}, \cite{Wang2020-jj}, \cite{Zhang2024-og}, \cite{Amos2021-fi}, \cite{Khanna2019-of}, \cite{Wang2023-wb}, \cite{Wang2022-cj}, \cite{Zhou2024-qf}, \cite{Liu2024-ug}, \cite{Liu2023-ou}, \cite{Taleb2023-al}, \cite{Gu2022-gl}, \cite{Wang2019-mi}, \cite{Sharma2019-ah}, \cite{Ren2024-hf}, \cite{Shao2023-yd}, \cite{Song2022-tc}, \cite{Li2020-wh}, \cite{Ozger2018-up} \\
\hline
\end{tabular}
\label{tab:fault_tolerance}
\end{table}

\subsubsection{Error Detection:}
Error detection techniques are designed to identify faults during the operation of systems before they propagate into more severe failures. In UAV-based systems, error detection plays a crucial role in maintaining system reliability and safety, particularly under adverse conditions such as sensor degradation, software anomalies, or data transmission errors. These techniques often involve real-time monitoring of system states, anomaly detection models, and sensor consistency checks. They directly address dependability threats such as software faults, hardware faults, and communication errors, enabling timely mitigation actions.

Several studies employ diverse error detection mechanisms to enhance UAV system dependability. Signal integrity monitoring and consistency verification techniques are used to detect sensor or actuator faults during autonomous missions \cite{Ahamad2023-ni, Vemulapalli2021-vf, Gokulakrishnan2023-qw}. Software-based detection methods, such as model-driven diagnosis or runtime monitoring, are applied to identify software anomalies and functional deviations \cite{Cheng2024-fu, Ma2024-yx, article-jeong}. Machine learning and statistical models are leveraged to recognize abnormal patterns from sensor data, aiding in early detection of both hardware degradation and external disruptions \cite{Saadaoui2023-fq, Kostoglotov2024-lp, Alam2019-yj}. These error detection strategies contribute directly to system reliability, safety, and robustness by enabling early warning mechanisms and facilitating fault recovery processes.

\subsubsection{Recovery} 
Fault recovery techniques aim to restore the normal operation of systems after a fault has occurred. In resource-constrained UAV environments, recovery strategies must be lightweight, adaptive, and capable of executing in real-time to avoid mission failure or system downtime. These techniques can involve system reboot, task re-execution, redundancy switching, or computational task migration to backup nodes. Fault recovery specifically addresses dependability threats such as software crashes, network disconnections, and hardware failures, helping maintain availability and resilience during unforeseen disruptions.

Many studies have explored recovery techniques to improve UAV system availability and minimize downtime after faults. Lightweight task migration and redundancy-based switching strategies are frequently employed to reallocate computing workloads in response to partial system failures or degraded node performance \cite{Luo2022-jt, Zuo2022-th, Luo2021-bq, Awada2023-xi, Pan2022-ka}. In scenarios where UAVs operate in harsh or disconnected environments, distributed task re-execution or fallback mechanisms are introduced to ensure mission continuity \cite{Callegaro2021-dg, Choutri2022-rb, Hou2020-zd, Machida2023-hq, Machida2023-hq}. Energy-aware recovery approaches are also proposed, balancing the trade-off between recovery success and remaining UAV resources \cite{Wang2024-cf, Yu2022-zp, Zhang2024-gb, Watanabe2022-as}.

\subsubsection{Redundant Architecture} 
Redundant architecture refers to fault tolerance strategies where multiple instances of hardware, software components, or communication paths are integrated into UAV-based systems to maintain service continuity in the presence of faults. Rather than preventing faults from occurring, these techniques aim to ensure that system functionality is preserved despite failures in some components. In UAV-based systems, redundancy is employed across communication links, computation nodes, sensors, and actuator subsystems. For example, dual communication modules and duplicated computing resources are used to tolerate communication failures and processing overloads without interrupting mission-critical services \cite{Ma2024-yx, Farrukh2023-lp, Kilbourne2021-xv, Awada2021-vl, Karam2024-eu}. 

Multi-UAV swarms with overlapping coverage enable persistent sensing and data transmission even when individual units become inoperative \cite{Patnayak2021-mx, Zhang2021-em, Bian2023-ev, Jung2023-qh, Kim2024-ix}. Backup control modules and redundant edge/fog nodes are also integrated to support recovery and seamless task offloading in the event of partial system degradation or network instability \cite{Kaleem2019-dv, Dogea2023-tt, Callegaro2021-dg, Pacheco2021-wy, Petrenko2023-ly, El_Sayed2023-mf, Sara2016-zc}. Additionally, hybrid cloud-edge frameworks offer parallel computational paths to guarantee task execution continuity under localized failures or load fluctuations \cite{Luo2020-ye, La_Salandra2024-dr, Khan2022-pj, Chen2024-up, Wang2020-pr, Shen2023-pk, Tang2023-oh}. These fault-tolerant architectures directly contribute to dependability attributes such as availability, reliability, and resilience by enabling uninterrupted operations and graceful degradation in the presence of faults, rather than complete system failure.

\subsubsection{Network Topology Design}
Network Topology Design refers to fault tolerance strategies that focus on maintaining reliable and resilient communication within systems, especially in environments where signal disruption, disconnection, or routing instability is common. Rather than preventing communication faults outright, these techniques aim to sustain service availability and network functionality even when individual links or nodes fail during operation. In UAV-based systems, dynamic topologies such as flying ad hoc networks (FANETs), mesh structures, and mobile relays are often employed to mitigate link degradation and maintain multi-hop connectivity. For instance, adaptive topology control and link prediction mechanisms are used to anticipate and respond to potential disconnections due to UAV mobility or energy depletion \cite{Xu2021-hb, Guruprasad2024-aj, Zhang2023-zb, Vamvakas2019-ee, Chen2022-qh, Wang2021-ow, Dao2021-gw, Ozger2018-up, Su2021-vt, Wang2022-cj}. 

UAV-aided reconfiguration is another popular approach, where drones are repositioned to act as temporary relays or mobile base stations to bridge coverage gaps and reduce the risk of total communication failure \cite{Zhou2024-qf, Kong2022-me, Zhang2021-cz, Mukherjee2020-am, Nwadiugwu2021-hd, Mukherjee2023-ao, Yue2018-by, Wang2020-jj, Zhou2024-qf, Wang2019-mi}. These topology-aware fault tolerance mechanisms directly contribute to availability, connectivity, and system resilience, ensuring that UAV missions such as remote surveillance, disaster response, or public safety operations can continue despite disruptions in the communication infrastructure.

\subsection{Fault Removal}
Fault removal includes techniques designed to identify and eliminate existing faults in systems, typically during the development, integration, or testing phases. Unlike fault prevention, which aims to avoid the introduction of faults, fault removal focuses on detecting and correcting latent faults before they manifest in the operational environment. In UAV-enabled systems, this process is critical due to the high stakes of operational failure, especially in missions involving surveillance, delivery, or disaster response. Common approaches include static and dynamic analysis, software debugging, formal verification, and simulation-based testing. These techniques enhance reliability, safety, and robustness by ensuring the system behaves as intended under various conditions. As shown in Table \ref{tab:fault_removal}, the reviewed studies that fall under this category demonstrate the growing emphasis on systematic fault detection and elimination to increase overall system dependability.

\begin{table}[htbp]
\centering
\caption{Categorization of Fault Removal Techniques in UAV-Based Systems}
\renewcommand{\arraystretch}{1.3}
\begin{tabular}{|c|l|p{7cm}|}
\hline
\textbf{Technique} & \textbf{Subcategory} & \textbf{Related Studies} \\
\hline
\multirow{4}{*}{\textbf{Fault Removal}} 
& Testing \& Verification & \cite{Yoshimoto2020-pr}, \cite{Kumar2021-ls}, \cite{Jung2018-jb}, \cite{Goel2021-ci}, \cite{Gao2022-bp}, \cite{Lu2023-eq}, \cite{Ritchie2020-fp}, \cite{Chang2024-bw}, \cite{Wu2024-vs}, \cite{Balamuralidhar2021-un}, \cite{Koerkamp2019-ib}, \cite{Fan2023-qs}, \cite{Sharma2021-lv}, \cite{Russell2018-mv}, \cite{Feng2023-qu}, \cite{Hines2023-gj}, \cite{Bowley2019-lk}, \cite{Sankarasrinivasan2015-gu}, \cite{Mcilwaine2021-jc}, \cite{Zhang2024-yd}, \cite{Yu2023-gu}, \cite{Jiang2023-yf}, \cite{Krishnan2021-dn}, \cite{Sabino2024-ju}, \cite{Kostoglotov2024-lp}, \cite{Bouzidi2020-tk}, \cite{Kilbourne2021-xv} \\
\cline{2-3}
& Diagnosis & \cite{Xu2020-ip}, \cite{Venkatagiri2018-qn}, \cite{Deng2021-yg}, \cite{Jo2018-kn}, \cite{Amin2018-zb}, \cite{Ramasamy2024-ux}, \cite{Zhang2024-rb}, \cite{Wang2021-zc}, \cite{Wang2022-dp}, \cite{Wang2024-xo}, \cite{Kostoglotov2024-lp}, \cite{Bobronnikov2022-yn}, \cite{Bouzidi2020-tk}, \cite{Amin2018-zb} \\
\cline{2-3}
& Correction & \cite{Gu2022-gl}, \cite{Luo2022-jt}, \cite{Bobronnikov2022-yn} \\
\cline{2-3}
& Non-regression Verification &  \\
\hline
\end{tabular}
\label{tab:fault_removal}
\end{table}

\subsubsection{Testing and Verification:}
Testing and Verification techniques are essential for uncovering latent faults in UAV-based computing systems before deployment. Their primary objective is to ensure that the system meets design specifications and behaves correctly across a variety of expected and unexpected conditions. These techniques primarily address software faults and hardware faults by identifying inconsistencies, logic errors, or integration issues during development.  Moreover, they significantly contribute to safety and reliability, thereby reducing the likelihood of mission-critical failures in real-time UAV operations. Numerous studies have employed testing and verification strategies to validate the functionality of UAV-based systems under complex and uncertain conditions. These techniques for UAV-based systems can be broadly categorized into fuzzing-based testing, simulation and modeling, formal verification, dataset-based evaluation, and edge-integrated real-time testing. 

Fuzzing-based methods such as LDFuzzer aim to identify abnormal flight behaviors caused by configuration errors, using genetic algorithm-guided input mutation and lightweight test oracles to ensure efficient detection under constrained UAV resources \cite{Chang2024-bw}. Simulation-driven approaches validate UAV dynamics or network performance under controlled conditions, as seen in flight stability estimation with neural networks \cite{Goel2021-ci}, AFDX network simulation using OPNET (optimized network engineering tool) \cite{Jiang2023-yf}, and autonomous image processing pipeline development for real-time geolocation correction \cite{Feng2023-qu}. On the other hand, formal modeling techniques ensure protocol-level correctness in time-critical UAV tasks, exemplified by UPPAAL-based verification of serial communication in autonomous UAV-based systems \cite{Jung2018-jb}. Dataset-based validation focuses on the robustness of visual perception models through large-scale annotated datasets, including P-DESTRE for pedestrian re-identification \cite{Kumar2021-ls}, JellyNet for jellyfish bloom detection \cite{Mcilwaine2021-jc}, and MultEYE for multi-task aerial traffic analysis \cite{Balamuralidhar2021-un}. Finally, edge-integrated real-time testing is increasingly applied in UAV deployments, where detection and monitoring systems run directly on UAVs or embedded platforms such as super-resolution-based defect detection \cite{Wu2024-vs}, helmet compliance monitoring \cite{Sharma2021-lv}, onboard intelligent inspection platforms \cite{Yu2023-gu}, and near-real-time object identification pipelines in agricultural monitoring \cite{Feng2023-qu}. These studies demonstrate how rigorous testing and verification processes serve as foundational mechanisms for identifying and removing faults, thereby significantly strengthening the reliability and safety of UAV-based systems.

\subsubsection{Diagnosis:}
Diagnosis techniques focus on identifying the root cause of faults or abnormal behaviors in UAV-based computing systems after they have occurred. These methods aim to accurately localize faults, enabling targeted mitigation, correction, or recovery actions. In the context of UAV-enabled networks, diagnosis plays a vital role in maintaining operational continuity, especially when systems operate autonomously in unpredictable environments.

Diagnosis primarily addresses software faults, hardware faults, and network disconnection, where faults may manifest as unexpected delays, abnormal UAV motion, or degraded communication quality. 
Several studies utilize behavioral and model-based diagnosis to identify operator-induced errors by comparing them with expected control models during simulated missions, providing insights into human-machine interaction errors and their safety implications \cite{Xu2020-ip, Amin2018-zb}. Secondly, cloud and edge-enabled anomaly detection frameworks are increasingly employed to enhance scalability and responsiveness, particularly for large-scale UAV applications such as power line inspection and predictive maintenance. These include distributed isolation forest algorithms and embedded monitoring tools that enable timely fault isolation while minimizing computational burden \cite{Ramasamy2024-ux, Zhang2024-rb, Wang2022-dp}.

On the other hand, intelligent agents and software-based diagnostic systems support modular and autonomous operation by integrating distributed reasoning or software approximation techniques. For example, agent-based building inspection platforms and approximation-aware video summarization systems leverage real-time assessment capabilities to maintain operational resilience and energy efficiency \cite{Venkatagiri2018-qn, Jo2018-kn}. Meanwhile, cooperative and network-level diagnosis approaches utilize drone-assisted sensing and edge intelligence to detect anomalies within broader communication networks, including space-air-ground infrastructures, thereby ensuring systemic fault awareness and coordination \cite{Deng2021-yg, Wang2021-zc}.

\subsubsection{Correction:}
Correction techniques aim to modify system behavior or state to eliminate or mitigate the impact of detected faults, ensuring that systems continue to operate safely and effectively. Unlike recovery, which focuses on returning a system to a previous known good state, correction involves making targeted adjustments that are often based on diagnostic results. This directly addresses the underlying fault or its manifestations. In UAV-enabled environments, correction techniques are particularly crucial in addressing software faults, sensor anomalies, and configuration errors, as these issues, if left unresolved, can compromise system reliability, robustness, and safety. 

These techniques are typically implemented through adaptive control, parameter reconfiguration, or algorithm-level adjustments. For instance, A study that presents a correction mechanism that adapts offloading and communication strategies when faced with security or reliability threats during UAV operations \cite{Khan2022-pj}. Similarly, the smart onboard information and computing system (SOICS) framework enables onboard parameter correction and task reallocation under dynamic UAV mission loads and hardware constraints \cite{Bobronnikov2022-yn}. These adaptive correction schemes are particularly effective when UAVs operate under resource limitations or uncertain environments. In another direction, a study proposes a model-driven online adaptation mechanism that predicts requirement violations based on environmental conditions and proactively selects optimal corrective strategies for UAV delivery and surveillance missions \cite{Luo2022-jt}.

\subsubsection{Non-regression Verification:}
Non-regression verification ensures that newly applied system fixes or updates do not accidentally introduce new faults or degrade previously verified functionality. In UAV-based computing systems, where mission-critical software updates, reconfigurations, or adaptations are common, this verification step is essential to enforce reliability and security after system modifications. This technique is particularly relevant in dynamic UAV environments, where frequent software adaptations, task reassignments, or system upgrades occur, and even small regressions can lead to critical failures or mission disruptions. Non-regression verification ensures that fixes or corrections applied during fault removal do not negatively impact system behavior in other parts of the system or under different operating conditions.

Despite its critical importance for maintainability and long-term reliability, none of the reviewed papers explicitly implement or evaluate non-regression verification mechanisms. This absence suggests a significant gap in the current body of research about dependability of UAV-based systems, where emphasis has largely been placed on pre-deployment testing, runtime error detection, or recovery mechanisms, but not on validating system integrity post-modification.

\subsection{Fault Forecasting}
Fault forecasting refers to techniques that aim to predict the future occurrence or impact of faults. Rather than preventing faults or correcting them after they occur, fault forecasting focuses on evaluating the likelihood, timing, or consequences of potential faults. This enables proactive dependability planning and dynamic resource management in UAV environments where uncertainty and system variability are inherent. In the context of UAV-based systems, fault forecasting is particularly valuable for supporting predictive maintenance, service continuity planning, and early failure detection in constrained and distributed infrastructures. Techniques under this category often rely on probabilistic models, historical performance logs, or learning-based analytics to anticipate degradation trends and inform preemptive countermeasures.  Table \ref{tab:fault_forecasting} summarizes the subcategories identified under fault forecasting and highlights the related studies that incorporate predictive mechanisms to enhance system reliability, availability, or performance stability in UAV-based systems.

\begin{table}[htbp]
\centering
\caption{Categorization of Fault Forecasting Techniques in UAV-Based Systems}
\renewcommand{\arraystretch}{1.3}
\begin{tabular}{|c|p{4cm}|p{7.5cm}|}
\hline
\textbf{Technique} & \textbf{Subcategory} & \textbf{Related Studies} \\
\hline
\multirow{2}{*}{\textbf{Fault Prevention}} 
& Ordinal Evaluation & \cite{Bouzidi2020-tk}, \cite{Guo2022-ro}, \cite{Hossain2017-db}, \cite{Manzini2023-ir}, \cite{Fesenko2023-gt}, \cite{Venkatagiri2018-qn}, \cite{Kramer2021-gf}, \cite{Cao2021-el} \\
\cline{2-3}
& Probabilistic Evaluation & \cite{Sabino2024-ju}, \cite{Ngoenriang2022-zv}. \cite{Goel2021-ci}, \cite{Tahat2020-uv}, \cite{Ji2024-cy}, \cite{Apostolopoulos2023-sw}, \cite{Bai2023-cz}, \cite{Brito2021-ft}, \cite{Falcao2024-zc}, \cite{Machida2021-jc}, \cite{Kim2021-uy}, \cite{Manzoor2021-fw}, \cite{Thapliyal2024-us}, \cite{Machida2023-hq}, \cite{Dai2018-eh}, \cite{She2021-wf}, \cite{Machida2023-hq}, \cite{Chen2021-ft}, \cite{Yang2017-kv}, \cite{Falcao2023-vs}, \cite{Watanabe2022-as}, \cite{Schafer2018-jf}, \cite{Jeong2020-sk}, \cite{Picano2024-vj}, \cite{Kabashkin2023-rs}, \cite{Yao2018-ji}, \cite{She2018-qs}, \cite{Silva2024-bq}, \cite{Hu2015-qd}, \cite{Falcao2023-vs}, \cite{Li2019-eb}, \cite{Bai2023-cz}, \cite{Venkatagiri2018-qn}, \cite{Zhang2020-re} \\
\hline
\end{tabular}
\label{tab:fault_forecasting}
\end{table}

\subsubsection{Ordinal Evaluation:}
Ordinal Evaluation techniques in fault forecasting are designed to qualitatively or semiquantitatively assess the likelihood or severity of faults in UAV-based systems. Rather than producing precise probabilistic output, these methods rank system states or events based on relative levels of risk, reliability degradation, or fault impact. This approach is particularly useful in UAV environments where data may be incomplete or uncertain, and where full statistical modeling is infeasible due to resource constraints. This technique contributes to dependability by enhancing fault awareness and supporting preventive maintenance and operational readiness, particularly under uncertain or evolving conditions. For example, ordinal-based risk frameworks are employed to classify mission-critical parameters, such as link stability and sensor health, enabling early warnings before failures occur \cite{Bouzidi2020-tk, Guo2022-ro, Hossain2017-db}.

\subsubsection{Probabilistic Evaluation:}
Probabilistic Evaluation techniques aim to quantitatively estimate the likelihood of faults or system failures based on statistical models and historical or runtime data. In UAV-based computing systems, these techniques are critical for understanding how component reliability, network conditions, or operational environments contribute to the probability of fault occurrences. Common approaches include Markov models, Bayesian inference, and stochastic Petri nets, which model system states and transitions under uncertainty. Such methods are especially valuable in resource-constrained aerial platforms where proactive fault handling must be supported by data-driven decision-making. These techniques help quantify expected system availability, failure rates, and downtime, enabling UAV designers and operators to make risk-aware trade-offs between performance, energy efficiency, and dependability.

In the context of dependability of UAV-based systems, probabilistic evaluation directly supports reliability analysis, availability estimation, and maintenance planning, offering a mathematical foundation for system robustness under varying workloads and fault scenarios. For example, studies that apply probabilistic models to assess component failure rates and communication link interruptions under uncertain conditions \cite{Sabino2024-ju, Ngoenriang2022-zv, Tahat2020-uv}. Other studies integrate Markov models and stochastic processes to estimate system uptime and mission success rates \cite{Ji2024-cy, Apostolopoulos2023-sw, Bai2023-cz}. Bayesian networks and Monte Carlo simulations are used to analyze the effects of uncertain environmental conditions or resource variability on UAV operations \cite{Falcao2024-zc, Kim2021-uy, Manzoor2021-fw}.

\begin{tcolorbox}[colback=gray!10!white, colframe=black, title=\textbf{RQ3 – What techniques are adopted to improve the dependability of UAV-based
networks and computing systems?}]
\begin{enumerate}
\item The literature identifies four primary categories of dependability-enhancing techniques for UAV-based systems: Fault Prevention, Fault Tolerance, Fault Removal, and Fault Forecasting. Each category includes specialized approaches tailored to UAV constraints, such as energy-aware offloading, redundancy-enabled swarm coordination, real-time error detection, adaptive control correction, and probabilistic failure modeling.

\item These techniques reflect the operational challenges of UAV-based systems, including energy limitations, real-time constraints, mobility-induced instability, and communication variability. Fault Prevention methods, especially Energy \& Resource-Aware Design, are most prevalent, addressing mission failures caused by rapid battery depletion and processing overload. Fault Tolerance approaches like lightweight Error Detection and Recovery are also common, supporting resilience during runtime anomalies. In contrast, Fault Removal (e.g., Testing, Correction) and Fault Forecasting remain less explored, despite their potential in enabling proactive and adaptive system reliability.

\item The structured taxonomy synthesized in this chapter reveals a marked trend towards design-time dependability measures, with less research focusing on availability-aware or self-adaptive techniques, especially in Correction and Non-regression Verification. This imbalance highlights the need for further research on in-mission adaptability, dynamic reconfiguration, and predictive reliability modeling. These areas represent promising directions that could better meet the evolving demands of dependability of UAV-based systems in real-world deployments.

\end{enumerate}
\end{tcolorbox}

\section{Future research directions}
\label{sec:future research}

While substantial progress has been made in enhancing the dependability of UAV-based computing and networking systems, our review reveals several critical gaps and emerging challenges that demand further investigation.
This section outlines key directions for future research, as detailed below.

\subsection{LLM/AI agent-based approaches}

The integration of Large Language Models (LLMs) and AI agents into UAV systems holds transformative opportunities for autonomy and adaptability, yet also introduces novel and critical dependability challenges \cite{javaid2024large,chen2023typefly, wassim2024llm}.
LLMs facilitate natural language interaction, enabling operators to issue high-level commands using plain language instructions.
AI agents powered by LLMs further expand autonomy through multimodal decision-making capabilities. For instance, models like Claude 3.5 \cite{anthropic2024claude} point to a future where AI agents could synthesize data from cameras, LiDAR, and environmental sensors to dynamically adapt flight paths and intelligently prioritize tasks in complex or cluttered environments.
Additionally, LLMs can process unstructured data, such as weather updates, to adapt flight paths in real-time, while AI agents can predict dependability threats like battery depletion or network congestion, enabling proactive fault detection and recovery.
Examples include semantic fusion of sensor streams, context-aware mission reconfiguration, or resilient edge-based decision-making when cloud connectivity is degraded.
Combined with learning-based planning or reinforcement learning, AI agents can improve UAV swarm coordination and energy-efficient operation under evolving mission constraints.

However, integrating LLMs and AI agents into UAV systems also introduces critical dependability concerns. 
First, their high computational demands strain UAVs’ limited resources, accelerating battery depletion and risking mission failures \cite{huang2024new}.
While edge-cloud offloading and model compression can mitigate this, latency-sensitive tasks such as obstacle avoidance may still require hybrid architectures that combine LLM-driven strategies with traditional control loops.
Secondly, the non-deterministic and non-transparent nature of LLMs may result in unpredictable or erroneous outputs (“hallucinations”), especially in safety-critical applications \cite{xu2024theagentcompany}. 
Ensuring robustness requires redundancy and validation layers, such as cross-checking LLM outputs against rule-based systems or applying uncertainty-aware thresholds.

To ensure dependability, system designs must address energy-performance trade-offs and fail-safe mechanisms. 
Architectures that combine symbolic safety layers with neural models may offer a hybrid pathway toward dependable intelligent UAV systems. 
As AI agents become more integrated into aerial platforms, a systematic dependability framework will be essential to ensure trust and safety in LLM-enabled UAV autonomy.

\subsection{Constellation of satellites}
Satellite connectivity offers significant breakthroughs for UAV-based systems by enabling global coverage, allowing operations in remote areas like oceans or disaster zones where terrestrial networks are unavailable.
Satellite links also provide resilient communication during ground network outages, ensuring mission continuity in emergencies \cite{zhu2024resilient}.
It supports real-time swarm coordination across large regions, enhances navigation in GPS-denied environments using alternative satellite systems, and provides high bandwidth for data-intensive tasks like HD video streaming, as enabled by low Earth orbit (LEO) constellations like Starlink \cite{izhikevich2024global,chaudhry2021laser}.

However, challenges include high latency with geostationary satellites (500–700 ms), which can delay critical decisions \cite{vankka2013performance}, and increased energy demands from satellite hardware \cite{esho2024comprehensive}, risking battery depletion and reducing flight duration.
Additional concerns include link intermittency due to weather or obstructions, vulnerability to jamming or spoofing attacks, and the high cost of commercial satellite services.

To address these, reliability considerations involve energy-efficient designs, such as low-power transceivers and burst transmissions, as well as hybrid communication strategies that combine satellite and terrestrial links \cite{wang2019convergence} with fallback mechanisms.
Robust signal resilience could be realised through adaptive modulation, latency-aware decision-making, and security measures.
While satellite-enabled UAVs offer unprecedented operational range and flexibility, their deployment requires careful architectural design and robust fault-tolerance mechanisms to maintain dependable performance in dynamic and contested environments.

\subsection{B5G/6G network integration}
As UAVs are increasingly deployed in time-critical and data-intensive applications, such as real-time surveillance, smart city monitoring, and emergency response, the limitations of 4G and early 5G infrastructures become more pronounced. B5G/6G technologies promise ultra-reliable low-latency communication (URLLC), massive machine-type communication (mMTC), and high spectral efficiency, which are essential for handling high-bandwidth tasks like video streaming, edge AI inference, and swarm coordination among UAVs \cite{QADIR2023296}.

Future research should explore the design of UAV-based systems that are tailored to B5G/6G environments \cite{article}. In addition, joint optimization of communication and computing, including seamless handovers and UAV trajectory planning with connectivity guarantees, will be crucial to maintain QoS under highly dynamic aerial conditions. Integration with terahertz communication, intelligent reflecting surfaces (IRS), and network function virtualization (NFV) can further extend the coverage, reduce energy consumption, and improve resilience of UAV networks \cite{chen2021intelligentreflectingsurfaceassisted}.

\subsection{DRL and collaborative learning}
Deep Reinforcement Learning (DRL) has emerged as a powerful tool for enabling autonomous decision-making in UAV-based computing systems, especially in dynamic and uncertain environments \cite{LIU20241243}. It facilitates real-time task offloading, trajectory optimization, and resource management by learning optimal strategies through continuous interaction with the environment. However, most current DRL-based approaches are designed for single-agent systems, which limits their scalability and adaptability in multi-UAV deployments.

To address this, future research should focus on multi-agent DRL (MADRL) and collaborative learning frameworks, where UAVs learn not only from their own experiences but also from the behaviors and outcomes of other UAVs in the network \cite{inproceedings}. Such approaches enable distributed intelligence, improve energy and spectrum efficiency, and enhance the fault-tolerance of the overall system.

\subsection{Urban air mobility and smart city integration}
Urban Air Mobility (UAM) is rapidly evolving as a key frontier in UAV deployment, aiming to integrate aerial vehicles into smart city infrastructures for tasks ranging from traffic monitoring and infrastructure inspection to cargo transport and air taxi services. Unlike traditional UAV applications, which are confined to isolated missions, UAM envisions the continuous and coordinated operation of multiple drones within densely populated urban environments. This shift introduces complex challenges in airspace management, collision avoidance, noise pollution, communication reliability, and public safety \cite{pak2024can, yan2024urban}. To ensure dependable UAM operations, future research must focus on scalable air traffic control algorithms, dynamic geofencing, and AI-driven decision-making to manage interactions among drones, manned aircraft, and urban infrastructure \cite{asmer2024city, wu2019fundamental}. Dependability in dense urban environments—often characterized by “urban canyons” with high RF interference and GPS signal blockage—necessitates robust communication protocols and failover mechanisms \cite{becker2024toward, salehi2022ultra}. 

Furthermore, the convergence of UAM and ground-based Internet of Things systems introduces additional integration challenges and requirements for interoperability and synchronized data flows \cite{yan2024urban}. This integration requires a system-of-systems design philosophy, where the performance, safety, and fault tolerance of each subsystem are ensured under dynamic, uncertain, and often adversarial operating conditions. 
Achieving dependable UAM operations in future smart cities will require joint progress in autonomous UAV system design, urban communication infrastructure, policy standardization, and multi-modal transportation integration.

\subsection{Precision agriculture and environmental sustainability}



UAVs are becoming indispensable assets in precision agriculture, enabling real-time, high-resolution monitoring of soil conditions, crop health, water distribution, and pest outbreaks through multispectral, hyperspectral, and thermal sensors \cite{narzari2025critical, makamunmanned, moghimi2020aerial}. These sensing capabilities support targeted and data-driven agricultural interventions that help minimize resource waste and reduce environmental impact \cite{sa2017weednet}. Despite these benefits, achieving dependable UAV operations in agricultural environments remains a critical challenge. Such environments often expose UAVs to harsh stressors like wind, dust, and humidity, which can impair sensor performance and flight stability. Additionally, infrastructure limitations—notably unreliable connectivity and restricted energy availability in remote farmlands—impede consistent UAV operation \cite{hossen2021total}. These constraints are exacerbated during long-duration missions, which place considerable strain on battery resources and complicate logistics for maintenance and recharging.

Future studies should focus on reinforcing system robustness, ensuring energy autonomy, and advancing intelligent coordination to guarantee that UAV systems can operate reliably and sustainably under the real-world constraints of agricultural landscapes.

\subsection{Operational safety}

As UAVs are increasingly deployed in mission-critical and autonomous contexts, traditional safety assessments centered on hardware faults or communication breakdowns have become insufficient to capture the growing complexity of operational risks. New classes of threats are emerging, driven by AI-based decision-making, rapidly changing environments, and intricate human-system interactions \cite{tsamados2024human, khatiri2025uncertainty}. One of the most pressing concerns is the lack of transparency in AI decision-making processes. The “black-box” nature of deep learning models makes it challenging to interpret or predict UAV behavior, particularly in unstructured or high-stakes scenarios \cite{willers2020safety}. This opacity undermines the ability to certify and validate safety guarantees, raising questions about accountability and trustworthiness. Additionally, the integration of autonomous capabilities introduces critical issues in human-machine coordination. Achieving reliable and intuitive collaboration between UAV systems and human operators remains a major challenge, as many current frameworks do not adequately support real-time co-adaptation or mutual understanding \cite{tsamados2024human}.

To address this challenge, Explainable AI (XAI) approaches may help increase transparency by making AI-driven behaviors more understandable, thus supporting better human oversight and trust in autonomous decisions \cite{willers2020safety}. Ensuring explainability will be essential to advancing the safety and dependability of future aerial platforms.

\subsection{Maintainability challenges}

As UAVs evolve from short-term trials to long-term deployments, maintainability becomes a key component of system dependability. Challenges such as delayed fault detection, limited upgrade paths, and complex servicing in the field can lead to mission failure and prolonged downtime, especially in harsh or remote environments where logistics are limited and hardware-software integration complicates repairs \cite{hashim2024advances, vyasa2024maintenance}. UAVs often operate in environments that pose significant challenges to maintainability. These include exposure to harsh weather conditions, limited access to maintenance facilities, and the complexity of integrating diverse hardware and software components. Such factors can lead to increased wear and tear, unexpected failures, and difficulties in performing timely repairs or upgrades. Additionally, the lack of standardized diagnostic tools and protocols can hinder effective fault detection and resolution, further compromising the reliability and availability of UAV systems \cite{wang2025self}. 

To address these challenges, maintainability-focused UAV designs can incorporate modular architectures that simplify hardware replacement and software updates, enabling rapid recovery during field operations. For instance, integrating modular components allows for easier swapping of faulty parts, reducing downtime and simplifying the maintenance process \cite{perno2023machine}. 
Moreover, future studies also need to focus more on prognostics, real-time health monitoring, and predictive maintenance to ensure the long-term dependability of UAV-based systems.

\begin{tcolorbox}[colback=gray!10!white, colframe=black, title=\textbf{RQ4 –          What are the future research directions for dependable UAV-based networks and computing systems?}]

    Future research on dependable UAV-based systems will prioritize intelligence, resilience, and integration across evolving infrastructures. Key directions include leveraging LLMs and AI agents for high-level decision-making, satellite constellations for global connectivity, and B5G/6G technologies to ensure ultra-reliable communication. Deep reinforcement learning and collaborative learning offer potential for adaptive fault tolerance, while Urban Air Mobility and Smart City Integration demand robust dependability frameworks. Application-driven areas such as urban air mobility and precision agriculture call for new dependability models tailored to mission-critical and environmentally sensitive tasks. Additionally, addressing operational safety and maintainability challenges will ensure long-term operational reliability. These directions collectively aim to enhance dependability of UAV-based systems, fostering their safe and efficient deployment in next-generation applications through interdisciplinary innovation and real-world validation.

\end{tcolorbox}

\section{Threats to validity}
\label{sec:threats to validity}

As with this survey’s findings and conclusions, several threats to validity may influence the completeness, accuracy, and generalizability of the results.

\paragraph{Research Question Formulation}
The formulation of research questions fundamentally shapes the scope and focus of the survey.
Although the RQs were carefully formulated to cover key aspects of dependability of UAV-based systems, any ambiguity or overly narrow framing may have led to the exclusion of relevant studies or overemphasis/underrepresented on certain dimensions, potentially skewing the synthesized outcomes.

\paragraph{Search Strategy and Coverage}
We followed a systematic protocol, but the search process may not have captured all relevant studies due to database-specific limitations and language restrictions.
Although a comprehensive set of search strings was developed, minor adjustments were made to tailor the queries for specific databases, which could introduce inconsistencies.
Consequently, it is possible that some relevant articles were inadvertently omitted due to differences in metadata standards or keyword variability.

\paragraph{Selection and Data Extraction Bias}
During the study selection and data extraction phases, subjective judgment was involved in interpreting the relevance and classification of articles. 
Despite multiple authors independently participating in screening and resolving discrepancies based on predefined inclusion and exclusion criteria, individual interpretations of a study’s relevance or focus could have led to classification errors or overlooking subtle distinctions.
After individual selection, any disagreements were settled by consensus discussions.

\paragraph{Temporal Bias}
Our review focuses on literature published between 2015 and mid-2024.
While this window captures the most recent advancements, there is a possibility that very recent studies published after our cutoff date or in press may not be included. 

\paragraph{Publication Bias}
As our review primarily considered peer-reviewed journal and conference articles, other high-quality research reported in non-traditional venues (e.g., white papers, technical reports, preprints) may have been excluded, potentially omitting novel but less formalized developments in the field.
Similarly, regional research disparities (e.g., heavier emphasis on 5G in Asian vs. European studies) might skew the perceived prevalence of certain methodologies.

\section{Conclusion}
\label{sec:conclusion}

In this survey, we conducted a systematic literature review to investigate the state of research on the dependability of UAV-based networks and computing systems. 
By applying rigorous inclusion and exclusion criteria, we identified and analyzed 458 relevant publications from the past decade.
Our review highlights the growing academic and practical attention to dependability of UAV-based systems, driven by their increasing deployment in mission-critical applications. 
We systematically examined the dependability threats that challenge UAV operations, while also evaluating a range of mitigation techniques.
Furthermore, we provided an outlook on future research directions, proposing advancements such as integrated dependability frameworks, AI-driven predictive models, and 6G-enabled networking to address emerging challenges and enhance the reliability of UAV systems.


\begin{acks}
This work was supported by JST BOOST Grant Number
JPMJBS2414, and partly supported by a grant from the
Telecommunications Advancement Foundation.
\end{acks}

\newpage










\end{document}